\input amstex
\documentstyle{amsppt}
\magnification=\magstephalf
 \addto\tenpoint{\baselineskip 15pt
  \abovedisplayskip18pt plus4.5pt minus9pt
  \belowdisplayskip\abovedisplayskip
  \abovedisplayshortskip0pt plus4.5pt
  \belowdisplayshortskip10.5pt plus4.5pt minus6pt}\tenpoint
\pagewidth{6.5truein} \pageheight{8.9truein}
\subheadskip\bigskipamount
\belowheadskip\bigskipamount
\aboveheadskip=3\bigskipamount
\catcode`\@=11
\def\output@{\shipout\vbox{%
 \ifrunheads@ \makeheadline \pagebody
       \else \pagebody \fi \makefootline 
 }%
 \advancepageno \ifnum\outputpenalty>-\@MM\else\dosupereject\fi}
\outer\def\subhead#1\endsubhead{\par\penaltyandskip@{-100}\subheadskip
  \noindent{\subheadfont@\ignorespaces#1\unskip\endgraf}\removelastskip
  \nobreak\medskip\noindent}
\def\endremark{\par\revert@envir\endremark\vskip\postdemoskip}
\outer\def\enddocument{\par
  \add@missing\endRefs
  \add@missing\endroster \add@missing\endproclaim
  \add@missing\enddefinition
  \add@missing\enddemo \add@missing\endremark \add@missing\endexample
 \ifmonograph@ 
 \else
 \vfill
 \nobreak
 \thetranslator@
 \count@\z@ \loop\ifnum\count@<\addresscount@\advance\count@\@ne
 \csname address\number\count@\endcsname
 \csname email\number\count@\endcsname
 \repeat
\fi
 \supereject\end}
\catcode`\@=\active
\CenteredTagsOnSplits
\NoBlackBoxes
\nologo
\def\today{\ifcase\month\or
 January\or February\or March\or April\or May\or June\or
 July\or August\or September\or October\or November\or December\fi
 \space\number\day, \number\year}
\define\({\left(}
\define\){\right)}
\define\Ahat{{\hat A}}

\define\CC{{\Bbb C}}

\define\Hom{\operatorname{Hom}}
\define\Map{\operatorname{Map}}

\define\RP{{\Bbb R\Bbb P}}
\define\RR{{\Bbb R}}

\define\ZZ{{\Bbb Z}}
\define\[{\left[}
\define\]{\right]}

\define\chiup{\raise.5ex\hbox{$\chi$}}
\define\cir{S^1}

\define\exertag #1#2{#2\ #1}

\define\inv{^{-1}}
\define\mstrut{^{\vphantom{1*\prime y}}}
\define\protag#1 #2{#2\ #1}

\define\res#1{\negmedspace\bigm|_{#1}}
\define\temsquare{\raise3.5pt\hbox{\boxed{ }}}

\define\theprotag#1 #2{#2~#1}

\define\xca#1{\removelastskip\medskip\noindent{\smc%
#1\unskip.}\enspace\ignorespaces }

\define\zmod#1{\ZZ/#1\ZZ}

\define\zt{\zmod2}

\define\rem#1{\marginalstar\begingroup\bf[{\eightpoint\smc{#1}}]\endgroup}
\def\strutdepth{\dp\strutbox} 
\def\marginalstar{\strut\vadjust{\kern-\strutdepth\specialstar}} 
\def\specialstar{\vtop to \strutdepth{ 
    \baselineskip\strutdepth 
    \vss\llap{$\bold{\Rightarrow}$ }\null}}

\NoRunningHeads

\input xy 
\xyoption{all}
\redefine\cir{S^1}
\loadbold
\loadeufm
\define\Alt{\operatorname{Alt}} 
\define\Bcls{\Cal{B}^{\text{cl}}_{t,\Sigma }}
\define\Bcl{\Cal{B}^{\text{cl}}}

\define\Bim{\operatorname{Bim}} 
\define\Bqtm{\Cal{B}^{\text{qtm}}}
\define\Bscl{\Cal{B}^{\text{scl}}}
\define\ER{E_{\RR}}
\define\ET{E_{\TT}}
\define\Ecls{\Cal{E}^{\text{cl}}_{t,\Sigma }}
\define\Ecl{\Cal{E}^{\text{cl}}}
\define\Eqtm{\Cal{E}^{\text{qtm}}}
\define\Escl{\Cal{E}^{\text{scl}}}
\define\Ext{\operatorname{Ext}}
\define\Fqtm{\Cal{F}^{\text{qtm}}}
\define\OZ{\Omega _\ZZ}
\define\Obb{\Omega _{\bb}}
\define\TT{\Bbb{T}}
\define\Tors{\operatorname{Tors}}
\define\VE{V_E}
\define\agp{\Cal{A}}
\define\bF{\check{F}}
\define\bb{\bar{b}}
\define\bul{^\bullet}
\define\cB{\check{B}}
\define\cE{\check{E}}
\define\cF{\check F}
\define\cG{\check G}
\define\cH{\check H}

\define\cKO{KO^\vee}
\define\cK{\check K}

\define\corr#1{\Ahat_E(#1)}
\define\cphi{\check \phi }
\define\crho{\check \imath }

\define\dd{\boldkey{d}}
\define\id{\operatorname{id}}

\define\pair{\{\cdot ,\cdot \}}
\define\pt{\text{pt}}
\define\sB{\Cal{B}}
\define\sC{\Cal{C}}
\define\sE{\Cal{E}}
\define\sG{\Cal{G}}
\define\sHY{\Cal{H}_Y}
\define\sH{\Cal{H}}
\define\sM{\Cal{M}}

\define\sY{\star}
\define\sZ{\Cal{Z}}
\define\stY{{s}_Y}
\define\tF{\tilde{F}}
\define\tom{\tilde{\omega}}
\define\tors{\text{torsion}}
\redefine\OE#1#2{\Omega _E(#1;\VE)^{#2}}

\refstyle{A}
\widestnumber\key{SSSSSSS}   
\document


	\topmatter
 \title\nofrills The Uncertainty of Fluxes \endtitle
 \author Daniel S. Freed\\Gregory W. Moore\\Graeme Segal  \endauthor
 \thanks The work of D.F. is supported in part by NSF grant DMS-0305505. The
work of G.M. is supported in part by DOE grant DE-FG02-96ER40949.  G.M.
would like to thank the Institute for Advanced Study for hospitality and the
Monell foundation for support during completion of this paper.  This research
was supported in part by the National Science Foundation under Grant
No\. PHY99-07949 to the Kavli Institute for Theoretical Physics.  This is
preprint NSF-KITP-05-119.  We also thank the Aspen Center for Physics for
providing a stimulating environment for discussions which led to this paper.
\endthanks
 \affil Department of Mathematics, University of Texas at Austin\\ Department
of Physics, Rutgers University\\ All Souls College, Oxford University\endaffil 
 \address Department of Mathematics, University of Texas, 1 University
Station, Austin, TX 78712-0257\endaddress 
 \email dafr\@math.utexas.edu \endemail
 \address Department of Physics, Rutgers University, Piscataway, NJ
08855-0849\endaddress 
 \email gmoore\@physics.rutgers.edu\endemail
 \address All Souls College, Oxford, OX1 4AL, UK \endaddress 
 \email graeme.segal\@all-souls.ox.ac.uk\endemail
 \date August 21, 2006\enddate
 \abstract In the ordinary quantum Maxwell theory of a free electromagnetic
field, formulated on a curved 3-manifold, we observe that magnetic and
electric fluxes cannot be simultaneously measured.  This uncertainty
principle reflects torsion: fluxes modulo torsion can be simultaneously
measured.  We also develop the Hamilton theory of self-dual fields, noting
that they are quantized by {\it Pontrjagin self-dual cohomology theories\/}
and that the quantum Hilbert space is $\zt$-graded, so typically contains
both bosonic and fermionic states.  Significantly, these ideas apply to the
Ramond-Ramond field in string theory, showing that its $K$-theory class
cannot be measured.
 \endabstract
	\endtopmatter

\document

Fluxes in the classical theory of electromagnetism and its generalizations
are real-valued and Poisson-commute.  Our main result is a Heisenberg
uncertainty principle in the quantum theory: magnetic and electric fluxes
cannot be measured simultaneously.  This observation applies to any abelian
gauge field, including the standard Maxwell field theory in four spacetime
dimensions as well as the $B$-field and Ramond-Ramond fields in string
theories.  It is the {\it torsion\/} part of the fluxes which experience
uncertainty---the nontrivial commutator of torsion fluxes is computed by the
{\it link pairing\/} on the cohomology of space, and there are always
nontrivial commutators if torsion is present.  We remark that torsion fluxes
arise from Dirac charge/flux quantization.  This Heisenberg uncertainty
relation goes against the conventional wisdom that the quantum Hilbert space
is simultaneously graded by the abelian group of magnetic and electric flux;
in fact, it is only graded by the free abelian group of fluxes modulo
torsion.  The most interesting example is the Ramond-Ramond field in
10-dimensional superstring theory.  Here the Dirac quantization law is
expressed in terms of topological $K$-theory, and conventional wisdom holds
that the quantum Hilbert space is graded by the integer $K$-theory of space.
The main result proves this wrong: the grading is only by $K$-theory modulo
torsion.  Notice that there are still operators reflecting the quantization
by the full $K$-theory group; the assertion is that these operators do not
all commute among themselves if there is torsion.

Our exposition in~\S{1} begins with the classical Maxwell equations.  We work
on a compact\footnote{There is also an uncertainty principle for fluxes if
space is noncompact; we hope to return to that topic in the future.}
3-dimensional smooth manifold~$Y$.  We first define the classical fluxes and
show that they Poisson commute.  The Hamiltonian formulation of Maxwell's
equations has Poisson brackets which are not invertible, so do not derive
from a symplectic structure, if the second cohomology of~$Y$ is nontrivial:
the symplectic leaves of the Poisson structure are parametrized by the
fluxes.  The most natural quantization of this system is as a family of
Hilbert spaces parametrized by the real vector space of fluxes.  Dirac charge
quantization is implemented in Maxwell theory by writing the electromagnetic
field as the curvature of a $\TT$-connection, where $\TT=U(1)$~is the circle
group.  There is now an action principle and the space of classical solutions
is a symplectic manifold, the tangent bundle to the space~$\sC(Y)$ of
equivalence classes of $\TT$-connections on~$Y$.  Its quantization is a
single Hilbert space, defined as the irreducible representation of the
Heisenberg group built from the product of~$\sC(Y)$ and its Pontrjagin dual.
(The salient features of Heisenberg groups and their representations are
reviewed in Appendix~A.)  Magnetic and electric fluxes are refined to take
values in the abelian group~$H^2(Y;\ZZ)$.  The Heisenberg uncertainty
relation, stated in \theprotag{1.20} {Theorem}, follows from the commutation
relations in the Heisenberg group.

Our second aim in this paper, carried out in~\S{2}, is to establish an
appropriate Hamiltonian quantization of generalized self-dual fields, such as
the Ramond-Ramond field in superstring theory.  We highlight the main issues
with the simplest self-dual field: the left-moving string on a circle, which
we simply call the self-dual scalar field.  Its quantization, which does not
quite follow from the usual general principles, serves as a model for the
general case.  The flux quantization condition for other gauge fields, both
self-dual and non-self-dual, is expressed in terms of a generalized
cohomology theory.  The fields themselves live in a {\it generalized
differential cohomology theory\/}.  We briefly summarize the salient points
of the differential theory.  The data we give in \theprotag{2.9}
{Definition} is sufficient for the Hamiltonian theory developed here; the
full Lagrangian theory requires a more refined starting point.  As in the
Maxwell theory one can write classical equations~\thetag{2.12} and a
Heisenberg group (\theprotag{2.14} {Theorem}), which now is $\zt$-graded.
The Hilbert space of the self-dual field is defined (up to noncanonical
isomorphism) as a $\zt$-graded representation of that graded Heisenberg
group.  The fact that the Hilbert space is $\zt$-graded, so in general has
fermionic states, is one of the novel points in this paper.  The
noncommutativity of quantum fluxes~\thetag{2.16} in the presence of torsion
is then a straightforward generalization of \theprotag{1.20} {Theorem} in the
Maxwell theory.  Section~2 concludes by showing how some common examples,
including the Ramond-Ramond fields, fit into our framework.

We call attention to one feature which emerged while investigating self-dual
fields in general.  For non-self-dual generalized abelian gauge fields {\it
any\/} generalized cohomology theory may be used to define the Dirac
quantization law.  However, for a self-dual field the cohomology theory must
itself be {\it Pontrjagin self-dual\/}.  See Appendix~B for an introduction
to generalized cohomology theories and duality.  Pontrjagin self-duality is a
strong restriction on a cohomology theory.  Ordinary cohomology, periodic
complex $K$-theory, and periodic real $K$-theory\footnote{$KO$-theory is
Pontrjagin self-dual with a shift; see \theprotag{B.11} {Proposition}.}  are
all Pontrjagin self-dual and all occur in physics as quantization laws for
self-dual fields.  Pontrjagin self-duality is not satisfied by most
cohomology theories.  For example, if the cohomology of a point in a
Pontrjagin self-dual theory contains nonzero elements in positive degrees,
then there are nonzero elements in negative degrees as well if the duality is
centered about degree zero.

In this paper we confine ourselves to the Hamiltonian point of view.  We only
construct the quantum Hilbert space and the operators which measure magnetic
and electric flux up to noncanonical isomorphism.  In future work we plan to
develop the entire Euclidean quantum field theory of gauge fields, both
self-dual and non-self-dual.  We emphasize that the Hamiltonian quantization
we use for a self-dual field is a special definition; it does not follow from
general principles of quantization---see the discussion at the beginning
of~\S{2}.  Perhaps one should not be surprised that the Hamiltonian theory
for self-dual fields requires a separate definition; after all, the same is
true for the Lagrangian theory~\cite{W}.  Our definition is motivated by some
preliminary calculations for a full theory of self-dual fields as well as by
the special case of the self-dual scalar field in two dimensions.

The role of Heisenberg groups and the noncommutativity of fluxes we find in
Maxwell theory was anticipated in~\cite{GRW}.

Our point of view in this paper is unapologetically mathematical.  The fields
under discussion are free, their quantum theory is mathematically rigorous,
whence our mathematical presentation.  In particular, we use the
representation theory of Heisenberg groups to define the quantum Hilbert
space of a free field.  Appendix~A reviews these ideas in the generality we
need.  A companion paper~\cite{FMS} presents our ideas and fleshes out the
examples in a more physical style.
 
We thank Dmitriy Belov, Jacques Distler, Sergey Gukov, Michael
Hopkins, Stephan Stolz, and Edward Witten for informative conversations.

 \head
 \S{1} Maxwell Theory
 \endhead
 \comment
 lasteqno 1@ 29
 \endcomment

 \subhead Classical Maxwell equations
 \endsubhead

Let $Y$ be a compact oriented\footnote{The orientation assumption is only for
convenience of exposition.  It is easily removed by working with differential
forms twisted by the orientation bundle.  The Hodge star operator maps
ordinary forms to twisted forms on a non-oriented manifold, so the electric
current~$j_E$ is a twisted form whereas the magnetic current~$j_B$ is
untwisted.} Riemannian 3-manifold and $M=\RR\times Y$ the associated
Lorentzian spacetime with signature~$(1,3)$.  The classical electromagnetic
field $F\in \Omega ^2(M)$ satisfies Maxwell's equations
  $$ \aligned
      dF&=j_B \\
      d*F&=j_E,\endaligned \tag{1.1} $$
where the magnetic and electric currents $j_B,j_E\in \Omega ^3(M)$ are
closed.  The Hodge star operator~$*$ is defined relative to the Lorentz
metric.  Let $\Sigma \subset Y$ be a closed oriented surface.  The {\it
classical magnetic flux\/} $\Bcls $ and {\it classical electric
flux\/}~$\Ecls$ are defined by
  $$ \Bcls(F) = \int_{\{t\}\times \Sigma }F,\qquad \qquad \Ecls(F)
     =\int_{\{t\}\times \Sigma }*F.  $$
It follows from~\thetag{1.1} that $\Bcls(F)$~and $\Ecls(F)$~are static---that
is, independent of time~$t$---if we impose~$j_B=j_E=0$.  Let $V$~be the
vector space of smooth solutions to the vacuum Maxwell equations
($j_B=j_E=0$) of finite energy.  Stokes' theorem implies that
both~$\Bcl_\Sigma (F)$ and $\Ecl_\Sigma (F)$~depend only on the homology
class of~$\Sigma $, so define functions
  $$ \Bcl,\Ecl\:H_2(Y)\times V \longrightarrow \RR \tag{1.2} $$
which are homomorphisms on $H_2(Y)\times \{F\}$ for all~$F\in V$.
Put differently, in this vacuum case $\Bcl,\Ecl$~are $H^2(Y;\RR)$-valued
functions on~$V$, simply the de Rham cohomology classes of~$F,*F$.

To express Maxwell's equations in Hamiltonian form, write
  $$ F=B-dt\wedge E \tag{1.3} $$
for $B=B(t)\in \Omega ^2(Y)$ and $E=E(t)\in \Omega ^1(Y)$; we set the speed
of light to one.  Let $\sY$~be the Hodge star operator on~$Y$ (relative to
its Riemannian metric).  Then $B$~and $\sY E$~are closed and their evolution
equations are
  $$ \frac{\partial B}{\partial t}= -d_YE,\qquad \frac{\partial (\star
     E)}{\partial t} = d_Y\star B.  \tag{1.4} $$
Let $W=\Omega ^2(Y)_{\text{closed}}\times \Omega ^2(Y)_{\text{closed}}$.  For
each time~$t$ there is an isomorphism~$V \to W$ obtained by evaluating the
restriction of~$(F,*F)$ to~$Y$ at time~$t$.  The Hamiltonian of the
electromagnetic field, its total energy, is
  $$ H = \frac 12\int_{Y}B\wedge \star B + E\wedge \star E.  $$
To put~\thetag{1.4} in Hamiltonian form we introduce a Poisson structure
on~$W$.

        \definition{\protag{1.5} {Definition}}
 A (translationally-invariant) {\it Poisson structure\/} on a real vector
space~$W$ is a skew-symmetric pairing
  $$ \pair\:W^*\times W^*\longrightarrow \RR  $$
on the space of linear functions on~$W$.  The {\it symplectic leaves\/} are
the affine translates in~$W$ of the annihilator of the kernel of~$\pair$. 
        \enddefinition

\flushpar
 If $\pair$~is nondegenerate, then $W$~is symplectic.  In general there is a
kernel, the subspace 
  $$ K = \{\ell \in W^*:\{\ell ,\ell '\}=0 \text{ for all~$\ell '\in W^*$}\},
     \tag{1.6} $$
and its annihilator is $\{w\in W:\ell (w)=0 \text{ for all $\ell \in K$}\}$.
The Poisson structure induces a Poisson bracket on polynomial functions
on~$W$.  The Poisson structure for Maxwell is most easily written in terms of
the linear functionals
  $$ \ell _\eta (B,\star E) = \int_{Y}\eta \wedge B ,\qquad \ell '_\eta
     (B,\star E) = \int_{Y} \eta \wedge \star E,\qquad \eta \in \frac{\Omega
     ^1(Y)}{d\Omega ^0(Y)}.  $$
Namely,
  $$ \{\ell \mstrut _{\eta _1},\ell \mstrut _{\eta _2}\} = \{\ell '_{\eta
     _1},\ell '_{\eta _2}\} = 0,\qquad
      \{\ell \mstrut _{\eta_1} ,\ell '_{\eta _2}\} = \int_{Y}d\eta _1\wedge
     \eta_2 .  $$
Then equation~\thetag{1.4} takes the Hamiltonian form
  $$ \frac{\partial B}{\partial t}= \{H,B\},\qquad \frac{\partial (\star
     E)}{\partial t} = \{H,\star E\}.  $$
The kernel~\thetag{1.6} of the Poisson bracket consists of~$\ell \mstrut
_\eta ,\ell '_\eta $ for {\it closed\/}~$\eta$, and so its annihilator is the
subspace
  $$ W_0 = d\Omega ^1(Y)\times d\Omega ^1(Y)\;\subset \;W.   $$
Therefore, the symplectic leaves are the fibers of the map~$\pi $ in 
  $$ 0 \longrightarrow W_0\longrightarrow W @>{\;\pi \;}>> H^2(Y;\RR)\times
     H^2(Y;\RR)\longrightarrow 0 \tag{1.7} $$
which assigns to~$(B,\star E)$ the pair~$\bigl([B]_{\text{dR}},[\star
E]_{\text{dR}} \bigr)$ of de Rham cohomology classes.

The classical fluxes~\thetag{1.2} may be viewed as $H^2(Y;\RR)$-valued linear
functions on~$W$, which together form the map~$\pi $ in~\thetag{1.7}.

        \proclaim{\protag{1.8} {Classical Fact}}
 The classical magnetic and electric fluxes~$\Bcl$ and~$\Ecl$ Poisson
commute.
        \endproclaim

\flushpar
 To make this precise, for a {\it closed\/} 1-form~$\eta $ on~$Y$ define the
linear functions $\Bcl(\eta ),\Ecl(\eta )\:W\to\RR$ by
  $$ \Bcl(\eta )(B,\star E) = \int_{Y}\eta \wedge B, \qquad \qquad
      \Ecl(\eta )(B,\star E) = \int_{Y}\eta \wedge \star E.  \tag{1.9} $$
Integration by parts shows 
  $$ \{\Bcl(\eta _1),\Ecl(\eta _2)\} = 0. $$

We pass to the quantum theory (without Dirac charge quantization) by
quantizing the affine symplectic fibers in~\thetag{1.7}, thus obtaining a
family of Hilbert spaces parametrized by~$H^2(Y;\RR)\times H^2(Y;\RR)$.  The
parameter is the pair of fluxes, which varies continuously over this real
vector space.  For simplicity we discuss only the quantization of the fiber
at~$(0,0)$, the symplectic vector space~$W_0$.  Briefly, one
writes~$W_0\otimes \CC$ as a direct sum of lagrangian subspaces and Hilbert
space completes the polynomial functions (Fock space) on one of the summands.
For finite dimensional symplectic vector spaces the resulting (projective)
Hilbert space~$\sH$ is independent of the lagrangian splitting.  For infinite
dimensional symplectic vector spaces one needs to fix a {\it polarization\/}
to specify the quantization.  In Hamiltonian field theory the natural
polarization is given by the energy operator~$i\,d/dt$: the complexification
of the space of classical solutions is the sum of positive energy solutions
and negative energy solutions.  The cleanest characterization of the quantum
(projective) Hilbert space~$\sH$ is as the unique irreducible representation
of associated Heisenberg group which is compatible with the polarization.  We
recall the definition of the Heisenberg group and leave further discussion to
Appendix~A.  Let $W_0$~be any symplectic vector space with symplectic
form~$\Omega $.  The Heisenberg group is a central extension of the
translation group~$W_0$ by the circle group~$\TT$ of unit complex numbers.
It is defined as the set $W_0\times \TT$ with multiplication
  $$ \bigl(w_1,\lambda _1\bigr)\cdot \bigl(w_2,\lambda _2\bigr) =
     \bigl(w_1+w_2\;,\; \lambda _1\lambda _2\exp(i\pi \Omega (w_1,w_2))
     \bigr). \tag{1.10} $$
The quantum Hilbert space~$\sH_0$ is heuristically the space of
$L^2$~functions on~$d\Omega ^1(Y)$, but rather than define a measure on this
infinite dimensional vector space we appeal to the representation theory of
the Heisenberg group.

 \subhead Semiclassical Maxwell theory: Dirac charge quantization
 \endsubhead

Our main concern is the modification of this discussion when {\it Dirac
charge quantization\/} is taken into account.  The quantization of charge
leads to the quantization of flux.  Thus on the spacetime ~$M=\RR\times Y$
the quantum magnetic and electric fluxes are constrained to live in a full
lattice inside the vector space~$H^2(Y;\RR)$, namely the image of integer
cohomology~$H^2(Y;\ZZ)$ in~$H^2(Y;\RR)$.  The map
  $$ H^2(Y;\ZZ)\longrightarrow  H^2(Y;\RR) \tag{1.11} $$
has a kernel, the torsion subgroup $\Tors H^2(Y;\ZZ)\subset H^2(Y;\ZZ)$, so
the lattice is naturally identified with~$H^2(Y;\ZZ)/\tors$.  A geometric
model which implements charge quantization takes the electromagnetic
field~$F$ to be $-1/2\pi i$~times the curvature of a connection~$A$ on a
principal circle bundle\footnote{In this Hamiltonian setup one postulates
that $P$~be the pullback of a bundle on~$Y$, i.e., time translation is lifted
to~$P$.}  $P\to M$.  This lifts the magnetic flux from $H^2(M;\ZZ)/\tors$ to
an element of the abelian group~$H^2(M;\ZZ)$, the Chern class of~$P$.  The
curvature of a connection depends only its isomorphism class.  The
space~$\sC(M)$ of isomorphism classes of smooth connections forms an infinite
dimensional abelian Lie group under tensor product of circle bundles with
connection.  The geometry of~$\sC(M)$ is important to us, so we pause to
elucidate it.
 
In this paragraph $M$~is any smooth manifold.  Let $\OZ^q(M),\;q>0,$ be the
set of closed $q$-forms on~$M$ with integral periods.  Also, let $\TT$~denote
the circle group.  Then
  $$ 0 \longrightarrow H^1(M;\TT) @>{\;\; i\;\; }>> \sC(M)@>\text{curvature}>>
     \OZ^2(M) \longrightarrow 0 \tag{1.12} $$
is an exact sequence of abelian groups.  Thus any closed 2-form with integral
periods is realized as the $\text{curvature}/(-2\pi i)$ of a connection and
the kernel of the curvature map is the group of isomorphism classes of flat
connections.  The latter is an abelian Lie group whose identity component is
the torus $H^1(M;\ZZ)\otimes \TT$ and whose group of components is the finite
group~$\Tors H^2(M;\ZZ)$.  This is encoded in the exact sequence
  $$ 1 \longrightarrow H^1(M;\ZZ)\otimes \TT \longrightarrow
     H^1(M;\TT) @>{\;\;\beta \;\;}>> \Tors H^2(M;\ZZ)\longrightarrow
     1, \tag{1.13} $$ 
where $\beta $~is the Bockstein homomorphism.  Also, $\OZ^2(M)$~is the union
of affine spaces~$\Obb$ of closed forms whose de Rham cohomology class is
$\bb\in H^2(M;\ZZ)/\tors$.  Let $\sC_{\bb}\subset \sC(M)$ be the preimage
of~$\Obb$; then $\sC_{\bb}\to\Obb$ is a principal bundle with structure
group~$H^1(M;\TT)$.  Another view of~$\sC(M)$ is exhibited by the exact
sequence
  $$ 0 \longrightarrow \Omega ^1(M)\bigm/\OZ^1(M) \longrightarrow \sC(M)
     @>\text{Chern}>> H^2(M;\ZZ) \longrightarrow 0; \tag{1.14} $$
any integral cohomology class of degree two is the Chern class of a principal
circle bundle and the kernel of the Chern class map is the set of connections
on the trivial bundle up to gauge equivalence.  Let $\sC_b$~denote the set of
equivalence classes of connections on a bundle with Chern class $b\in
H^2(M;\ZZ)$ and $\bb$~the image of~$b$ in de Rham cohomology.  Then
$\sC_b\to\Obb$ is a principal bundle with structure group the torus
$H^1(M;\ZZ)\otimes \TT$.  The group~$\sC(M)$ is naturally a Lie
group.  The quotient~$H^2(M;\ZZ)$ in~\thetag{1.14} is its group of components
and the kernel in that exact sequence is its identity component.

There is an action principle for Maxwell theory on the space\footnote{In
fact, the fields do not form a space but rather the groupoid of
$\TT$~connections on~$M$.  The lagrangian is gauge-invariant, so determines a
function on the space~$\sC(M)$ of equivalence classes.} of fields~$\sC(M)$.
Namely, 
  $$ L(A) = -\frac 12 F_A\wedge *F_A,  $$
where $-2\pi iF_A$~is the curvature of the connection~$A$.  Cauchy data at
fixed time identifies the space~$\sM$ of solutions to the corresponding
Euler-Lagrange equations as the tangent bundle to~$\sC(Y)$.  Now, in contrast
to the classical theory considered above, the entire space~$\sM$ has a
symplectic structure.  The magnetic and electric fluxes are defined as
in~\thetag{1.9}: for a closed 1-form~$\eta $ on~$Y$ we have the functions
$\Bscl(\eta ),\Escl(\eta )\:\sM\to \RR$ given by
  $$ \Bscl(\eta )(A) = \int_{Y}\eta \wedge F_A, \qquad \qquad
      \Escl(\eta )(A) = \int_{Y}\eta \wedge *F_A,
       $$
where the integrals are computed at any fixed time.  The semiclassical
magnetic flux is quantized---if $\eta $~represents an integral class then
$\Bscl$~is integer-valued---whereas $\Escl$~is not.

        \proclaim{\protag{1.15} {Semiclassical Fact}}
 The semiclassical magnetic and electric fluxes Poisson commute: 
  $$ \bigl\{\Bscl(\eta _1),\Escl(\eta _2)\bigr\}=0.  $$
        \endproclaim

\flushpar
 In fact, $\Bscl(\eta _1)$~is locally constant on~$\sM$, since
$[F_A]_{\text{dR}}$ takes discrete values, so $\Bscl(\eta _1)$~Poisson
commutes with {\it any\/} function on~$\sM$.

To motivate our definition of the quantum electric flux below we observe that
any function on a symplectic manifold generates an infinitesimal symplectic
diffeomorphism.  A linear function on a symplectic vector space generates an
infinitesimal translation, and for~$\Escl(\eta )$ it is infinitesimal
translation by~$\eta $, viewed as a static connection on the trivial bundle.
Its equivalence class in~$\sC(Y)$ lies in the torus ~$H^1(Y;\ZZ)\otimes \TT$.
In the quantum version below this torus is augmented to the compact abelian
Lie group~$H^1(Y;\TT)$.

 \subhead Quantum Maxwell theory
 \endsubhead

We quantize the semiclassical Maxwell theory.  Recalling that the space~$\sM$
of classical solutions may be identified with the tangent bundle to~$\sC(Y)$,
we see that the Hilbert space~$\sHY$ is {\it heuristically\/} the space of
$L^2$~ functions on~$\sC(Y)$.  It is naturally graded by the
group~$H^2(Y;\ZZ)$ of components of~$\sC(Y)$; the homogeneous subspaces
consist of $L^2$~functions supported on a single component:
  $$ \sHY = \bigoplus\limits_{b\in H^2(Y;\ZZ)}\sH^b. \tag{1.16} $$
 This is the grading by magnetic flux.  As for electric flux, observe that
the group of flat connections~$H^1(Y;\TT)$ acts on~$\sC(Y)$ by tensor
product.  This induces a representation of~$H^1(Y;\TT)$ on $L^2$~functions,
and we decompose~$\sHY$ according to the group of characters\footnote{The
identification of the characters with~$H^2(Y;\ZZ)$ uses Poincar\'e duality,
which is an important ingredient in the general picture.}
$\Hom\bigl(H^1(Y;\TT),\TT \bigr)\cong H^2(Y;\ZZ)$:
  $$ \sHY = \bigoplus\limits_{e\in H^2(Y;\ZZ)}\sH_e.  \tag{1.17} $$

        \proclaim{\protag{1.18} {Main Observation}}
 The gradings~\thetag{1.16} and~\thetag{1.17} do not necessarily induce a
simultaneous grading of~$\sHY$ by magnetic and electric flux.
        \endproclaim

\flushpar 
 In this heuristic picture, where $\sHY$~is the space of $L^2$~functions
on~$\sC(Y)$, the main observation follows immediately from the fact that
translation by an element of~$H^1(Y;\TT)$ does not preserve the components
of~$\sC(Y)$, which are labeled by~$H^2(Y;\ZZ)$.  In fact, the action on the
group of components is by translation via the Bockstein homomorphism~$\beta $
in~\thetag{1.13}.  In other words, the identity component of~$H^1(Y;\TT)$
{\it does\/} preserve the group of components, and the noncommutativity is
measured strictly by the torsion subgroup~$\Tors H^2(Y;\ZZ)$ of the second
cohomology.

        \proclaim{\protag{1.19} {Amelioration}}
 The Hilbert space~$\sHY$ is simultaneously graded by magnetic and electric
fluxes modulo torsion, that is, by the abelian group $H^2(Y;\ZZ)/\tors \times
H^2(Y;\ZZ)/\tors $.
        \endproclaim

\flushpar
 The main observation is in force whenever $Y$~has torsion in its second
cohomology, or equivalently in its first homology.  For example, take $Y$~to
be a three-dimensional lens space, such as real projective space~$\RP^3$.

 These observations can be formulated more sharply.  For each~$\omega \in
H^1(Y;\TT)$ there are (scalar-valued) linear operators~$\Bqtm(\omega
),\Eqtm(\omega )$.  The operator~$\Bqtm(\omega )$ is multiplication
by~$\langle b,\omega \rangle$ on~$\sH^b$; the operators~$\Eqtm(\omega )$ form
the representation of $H^1(Y;\TT)$ on~$\sHY$.

        \proclaim{\protag{1.20} {Theorem}}
 For~$\omega _1,\omega _2\in H^1(Y;\TT)$ we have
  $$ [\,\Bqtm(\omega _1),\Eqtm(\omega _2)\,] = (\omega _1\smile\beta\omega
     _2)[Y]\,\id_{\sHY}, \tag{1.21} $$
where $\beta $~is the Bockstein in~\thetag{1.13} and $[Y]$ the fundamental
class of~$Y$ in homology.
        \endproclaim
 
\flushpar
 The left hand side of~\thetag{1.21} is the {\it group\/} commutator
$\Bqtm(\omega _1)\Eqtm(\omega _2)\Bqtm(\omega _1)\inv \Eqtm(\omega _2)\inv $.
The pairing on the right-hand side of~\thetag{1.21} is symmetric and depends
only on ~$\beta \omega _1,\beta \omega _2$, so factors to a symmetric pairing
  $$ \tau \:\Tors H^2(Y;\ZZ)\times \Tors H^2(Y;\ZZ)\longrightarrow
     \TT, \tag{1.22} $$
the so-called {\it link pairing\/} or {\it torsion pairing\/} in cohomology.
Poincar\'e duality implies that $\tau $~is a perfect pairing: $\Tors
H^2(Y;\ZZ)$ is its own Pontrjagin dual.

These heuristics are made rigorous, and \theprotag{1.20} {Theorem} is proved,
by defining the quantum Hilbert space~$\sHY$ as a representation of a
generalized Heisenberg group.  (See Appendix~A for more details.)

        \definition{\protag{1.23} {Definition}}
 Let $\agp$~be an abelian group and $\psi \:\agp\times \agp\to\TT$ a
{\it 2-cocycle\/}, that is, 
  $$ \psi (a_1,a_2)\psi (a_1+a_2,a_3) = \psi (a_1,a_2+a_3)\psi
     (a_2,a_3),\qquad a_1,a_2,a_3\in \agp. \tag{1.24} $$
The group~$\sG(\agp,\psi )$ attached to~$(\agp,\psi )$ is the set~$\agp\times
\TT$ with multiplication
  $$ \bigl(a_1+a_2\;,\;\lambda _1\lambda _2\,\psi (a_1,a_2)
     \bigr).  $$
Commutators in~$\sG(\agp,\psi )$ are measured by a map $s\:\agp\times
\agp\to\TT$: 
  $$ \bigl[ (a_1,\lambda _1),(a_2,\lambda _2)\bigr]= \bigl[0,s(a_1,a_2)\bigr]
     = \bigl[0,\psi (a_1,a_2)\; \psi (a_2,a_1)\inv \bigr],\qquad a_1,a_2\in
     \agp.\tag{1.25} $$
The map~$s$ is {\it bimultiplicative\/}---a homomorphism in each variable
separately---and is {\it alternating\/}: $s(a,a)=1$ for all~$a\in \agp$.  We
say $s$~is {\it nondegenerate\/} if for all~$a_1\in \agp$ there
exists~$a_2\in \agp$ such that $s(a_1,a_2)\not= 1$.  If $s$~is nondegenerate,
then we call~$\sG(\agp,\psi )$ a {\it Heisenberg group\/}.
        \enddefinition

\flushpar
 If $\agp$~is a Lie group and $\psi $~is smooth, then $\sG(\agp,\psi )$~ is
also a Lie group.  The Heisenberg group associated to a symplectic vector
space~$(V,\Omega )$ is~$\sG(V,e^{i\pi \Omega})$; see~\thetag{1.10}.  The
Heisenberg group is a central extension of~$\agp$ by the circle group:
  $$ 1 \longrightarrow \TT \longrightarrow \sG(\agp,\psi )\longrightarrow
     \agp\longrightarrow 0.  $$
 
For quantum Maxwell theory we take 
  $$ \agp=\sC(Y)\times \sC(Y). \tag{1.26} $$
The pairing~$\psi $ is defined in terms of a symmetric pairing
  $$ \sigma \:\sC(Y)\times \sC(Y)\longrightarrow \TT  $$
on connections, namely
  $$ \psi \bigl((A_1,A_2),(A'_1,A'_2) \bigr) = \sigma
     (A_1,A'_2). \tag{1.27} $$
Notice that $\psi $~vanishes on~$H\times H$ for the subgroups~$H=\sC(Y)\times
\{0\}$ and~$H=\{0\}\times \sC(Y)$, so the Heisenberg group is canonically split
over these subgroups.  The simplest way to define~$\sigma $ is to use the
topological fact that every compact oriented 3-manifold~$Y$ with a pair of
circle connections~$A_1,A_2$ bounds a compact oriented 4-manifold~$X$ over
which the connections extend.  Let $-2\pi i\tF_j,\,j=1,2$, denote the
curvatures of the extended connections.  Then
  $$ \sigma (A_1,A_2) = \exp\bigl(2\pi i\int_{X} \tF_1\wedge \tF_2
     \bigr). \tag{1.28} $$
We remark that the diagonal value~$\sigma (A,A)$ is the Chern-Simons invariant
of the connection~$A$.  For $\alpha _1\in \Omega ^1(Y)/\OZ^1(Y)$
(see~\thetag{1.14}) this formula simplifies to 
  $$ \sigma (\alpha _1,A_2) = \exp\bigl(2\pi i\int_{Y} \alpha _1\wedge
     F_2\bigr),  $$
where $-2\pi iF_2$~is the curvature of the connection~$A_2$.  If $\omega
_1$~and $\omega _2$~are flat connections, i.e., live in the
subgroup~$H^1(Y;\TT)$ (see~\thetag{1.12}), then
  $$ \sigma (\omega _1,\omega _2) = (\omega _1\smile\beta \omega _2)[Y]
      $$
is the link pairing $\tau (\beta \omega _1,\beta \omega _2)$
(see~\thetag{1.22}).
 
\theprotag{A.5} {Proposition} in the Appendix~A asserts that the Heisenberg
group $\sG(\agp,\psi )$ has a unique irreducible unitary representation up to
isomorphism which is compatible with positive energy and on which the central
circle acts by scalar multiplication.

        \definition{\protag{1.29} {Definition}}
 The quantum Hilbert space~$\sHY$ of Maxwell theory is this unique irreducible
representation of~$\sG(\agp,\psi )$. 
        \enddefinition

\flushpar
 Recall~\thetag{1.27} that $\sG(\agp,\psi )$~is canonically split over
$\sC(Y)\times \{0\}$ and~$\{0\}\times \sC(Y)$.  The operators~$\Bqtm(\omega
)$ and~$\Eqtm(\omega )$ for~$\omega \in H^1(Y;\TT)$ are defined by
restricting the representation in \theprotag{1.29} {Definition} to the lifts
of the subgroups~$H^1(Y;\TT)\times\nobreak \{0\}$ and~$\{0\}\times
H^1(Y;\TT)$ of~$\sC(Y)\times \sC(Y)$ in~$\sG(\agp,\psi )$.  \theprotag{1.20}
{Theorem}, and so our main observation, now follows directly from the
commutation relations of the Heisenberg group.

 \head
 \S{2} Self-dual abelian gauge fields
 \endhead
 \comment
 lasteqno 2@ 28
 \endcomment

The simplest self-dual field is the self-dual scalar field in 2-dimensional
field theory, which with Dirac charge quantization takes values in the
circle~$\TT$.  It is also called a left-moving string.  Its motion is
described by a $\TT$-valued function of time~$t$ and space~$x$ of the form
$f(t,x)=\phi (x+t)$, which is a general solution to the first-order wave
equation
  $$ \partial f/\partial t = \partial f/\partial x. \tag{2.1} $$
When space is a circle the classical motions are parametrized by the loop
group~$L\TT = \Map(\cir,\TT)$.  Now any~$\phi \in L\TT$ may be uniquely
decomposed as
  $$ \phi (x) = \lambda e^{iwx}\exp\bigl(i\sum\limits_{n\not= 0}q_ne^{inx}
     \bigr),  $$
where $w\in \ZZ$~is the winding number, $\lambda \in \TT$, and
$q_{-n}=\overline{q_n}$ are complex numbers.  This defines a decomposition of
the loop group 
  $$ L\TT \cong \TT\times \ZZ\times V \tag{2.2} $$
as the product of the circle group, the integers, and a real vector space.
In physics $V$~is called the space of ``oscillators''.  There is a
Hamiltonian description of the motion.  The Poisson brackets define a
nondegenerate pairing on~$V$, but the overall structure is degenerate: the
symplectic leaves are parametrized by~$\TT\times \ZZ$.  The quantization of
this system does not follow standard rules, which would give a family of
Hilbert spaces parametrized by~$\TT\times \ZZ$.  Rather, we observe that
$\TT$~and $\ZZ$~are Pontrjagin dual, and in fact the entire loop
group~\thetag{2.2} is Pontrjagin self-dual.  So there is a Heisenberg central
extension, the standard central extension of the loop group at level one.
Furthermore, we introduce a $\zt$-grading.  

        \definition{\protag{2.3} {Definition}}
 A {\it $\zt$-grading\/} on a topological group~$\sG$ is a continuous
homomorphism $\epsilon \:\sG\to\zt$.  A {\it graded representation\/} of a
$\zt$-graded group is a $\zt$-graded Hilbert space $\sH^0\oplus \sH^1$ and a
homomorphism $\sG\to GL(\sH^0\oplus \sH^1)$ such that even elements of~$\sG$
preserve the grading on~$\sH^0\oplus \sH^1$ and odd elements reverse it.
        \enddefinition

\flushpar
 A $\zt$-grading~$\epsilon $ is constant on the identity component, so for a
Lie group does not induce any structure on the Lie algebra.  In particular,
there are no sign rules for a grading on a group.  For the loop group~$L\TT$
there is a unique nontrivial grading according to the parity of the winding
number, and it lifts to a grading of the Heisenberg central extension.  The
unique irreducible unitary representation of the Heisenberg central extension
is graded---see the discussion at the end of Appendix~A---and is defined to
be the quantum Hilbert space of the self-dual scalar.  That the Hilbert space
is $\zt$-graded is expected from the Bose-Fermi correspondence in two
dimensions.  Another motivation for the grading is~\cite{S,(12.3)}.

More complicated self-dual fields, such as the Ramond-Ramond fields of
superstring theory, lead to a Pontrjagin self-dual abelian group which
generalizes~\thetag{2.2}.  Its definition requires new ideas.  The Dirac
quantization law is implemented by a cohomology theory; for Ramond-Ramond
fields it is a flavor of $K$-theory.  The abelian group which plays the role
of~\thetag{2.2} is then a {\it differential\/} cohomology group built by
combining the cohomology theory with differential forms.  For ordinary
cohomology these groups were first introduced by Cheeger-Simons~\cite{CS} and
are a smooth version of Deligne cohomology~\cite{D}.  The generalization we
need is developed by Hopkins-Singer~\cite{HS}.\footnote{To define gauge
fields and gauge transformations one needs to introduce groupoids which
represent the differential cohomology groups.  Several models are considered
in~\cite{HS}.  For the Hamiltonian theory as discussed here we need only the
differential cohomology groups of gauge equivalence classes.}  We summarize
what we need and then go on to define self-dual fields and their quantum
Hilbert space.  Non self-dual fields quantized by ordinary cohomology, such
as the $B$-field in Type~II superstring theory, fit into our theory by
doubling and considering them as self-dual (\theprotag{2.18} {Example});
there is an equivalent direct treatment as a non-self-dual field.

 \subhead Generalized differential cohomology	
 \endsubhead

Let $E$~be a multiplicative cohomology theory, and $\ER,\ET$~the associated
theories with real and circle coefficients.  (See Appendix~B for a brief
introduction.) The real $E$-cohomology of a point~$\VE\bul=E\bul(\pt; \RR)$
is a $\ZZ$-graded real vector space.  For any space~$X$ there is a natural
map
  $$ E\bul(X)\longrightarrow H(X;\VE)\bul \tag{2.4} $$
whose image is a full lattice and whose kernel is the torsion subgroup.  The
codomain---the space on the right hand side of~\thetag{2.4}---is a direct sum
of ordinary real cohomology groups.  The total degree is the sum of the
cohomological degree and the degree in~$\VE$.  In case $E$~is ordinary
cohomology, $V_H\bul$~equals~$\RR$ in degree zero and vanishes otherwise, and
\thetag{2.4} is the map~\thetag{1.11}.  In case $E$~is complex $K$-theory,
$V_K\bul\cong \RR[u,u\inv ]$ is a Laurent series ring with the inverse
Bott element~$u$ of degree~2.  Now suppose $M$~is a smooth manifold.  The
generalized differential cohomology~$\cE\bul(M)$ is a $\ZZ$-graded abelian
Lie group.  We content ourselves here with a description of its geometry,
analogous to the discussion of the geometry of~$\sC(M)=\check H^2(M)$
in~\S{1}.  First, the $E$-cohomology with circle coefficients is a compact
abelian group whose identity component is a torus, as indicated in the
sequence
  $$ 1 \longrightarrow E^{d-1}(M)\otimes \TT \longrightarrow
     \ET^{d-1}(M) @>{\;\;\beta \;\;}>> \Tors E^d(M)\longrightarrow 1,
      $$
valid for any~$d\in \ZZ$.  (The $E$-cohomology is graded by the integers and
can be nonzero in negative degrees, as for example for $K$-theory.)  Let
$\OE{M}{q}$~denote the space of closed $\VE$-valued differential forms of
total degree~$q$ whose de Rham cohomology class lies in the image
of~\thetag{2.4}.  There are exact sequences
  $$ 0 \longrightarrow \ET^{d-1}(M) @>{\;\;i\;\;}>> \cE^d(M)@>\text{field
     strength}>> \OE Md \longrightarrow 0 \tag{2.5} $$
and 
  $$ 0 \longrightarrow \Omega (M;\VE)^{d-1}\bigm/\OE M{d-1} \longrightarrow
     \cE^d(M) @>\text{characteristic class}>> E^d(M) \longrightarrow
     0. \tag{2.6} $$
In the physics language, a gauge field of degree~$d$ up to gauge equivalence
is an element of $\cE^d(M)$.  Its field strength is computed using the
labeled map in~\thetag{2.5}; the isomorphism classes of flat fields form the
compact abelian group which is its kernel.  A gauge field also has a magnetic
flux, the characteristic class in~\thetag{2.6}, and the space of equivalence
classes of gauge fields with trivial magnetic flux is a quotient space of
differential forms.
 
Because $E$~is a {\it multiplicative\/} cohomology theory, the associated
differential theory has a graded multiplication
  $$ \cE^{d}(M)\otimes \cE^{d'}(M)\longrightarrow \cE^{d+d'}(M).  $$
It combines the wedge product on differential forms and the multiplication
in~$E$: the field strength of the product is the wedge product of the field
strengths and the characteristic class of the product is the $E$-product of
the characteristic classes.  Also, if $\omega \in \ET^{d-1}(M)$ and $\cF\in
\cE^{d'}(M)$, then the product~$i(\omega )\cdot \cF$ only depends on the
characteristic class of~$\cF$ and is the usual cohomological product
  $$ \ET^{d-1}(M)\otimes E^{d'}(M)\longrightarrow
     \ET^{d+d'-1}(M)\longrightarrow  \cE^{d+d'}(M).  $$
In particular,
  $$ i(\omega )\cdot i(\omega ') = \omega\cdot  \beta \omega '=\tau (\beta
     \omega ,\beta \omega '),\qquad \omega \in \ET^{d-1}(M),\quad \omega
     '\in \ET^{d'-1}(M),  $$
where $\tau $~is the ``link (torsion) pairing''
  $$ \tau \:\Tors E^{d}(M)\otimes \Tors E^{d'}(M)\longrightarrow
     \Tors \ET^{d+d'-1}(M). \tag{2.7} $$
Similarly, if $\alpha $~lies in the kernel of~\thetag{2.6} then the
product~$\alpha \cdot \cF$ only depends on the field strength of~$\cF$ and is
given by wedge product.

Integration (also called pushforward or direct image) is defined in
generalized differential cohomology.  Suppose $M$~is a compact $n$-manifold
which is oriented for $E$-cohomology.  An orientation for differential
$\cE$-cohomology includes an orientation for $E$-cohomology, but might
involve more data.  For ordinary cohomology an $\check H$-orientation is an
$H$-orientation is the usual notion of orientation.  For real $KO$-theory a
$KO$-orientation is a spin structure on~$M$.  A differential
$\cKO$-orientation is a spin structure together with a Riemannian metric.
Similarly, for complex $K$-theory a $K$-orientation is a $\text{spin}^c$
structure, whereas a differential $\check K$-orientation is a
$\text{spin}^c$-structure together with a Riemannian metric and compatible
covariant derivative on the $\text{spin}^c$ structure.  For a compact
$\cE$-oriented $n$-manifold~$M$ we have the integration map
  $$ \int_{M}\:\cE\bul(M)\longrightarrow \cE^{\bullet-n}(\pt).  $$
There is an extension to integration in fiber bundles and for arbitrary maps
as well.  These integrations satisfy the usual Stokes' theorem and are
compatible with the corresponding integration $E$-cohomology.  However, in
general it does not commute with the field strength: there is an invertible
differential form~$\corr M\in \Omega (M;\VE)^0$ such that for~$\cF\in
\cE^{\bul}(M)$ the field strength of~$\int_{M}\cF$ is
  $$ \int_{M} \corr M\wedge F, \tag{2.8} $$
where $F$~is the field strength of~$\cF$.  This form is derived from the
Riemann-Roch theorem, which relates the pushforwards in~$E$ and ordinary
cohomology.  Thus when $E$~is ordinary cohomology this form is~1; its value
for real $K$-theory explains our choice of notation.  There is an extension
to integration in fiber bundles.  Finally, if $\cF=i(\omega )$
in~\thetag{2.5}, then $\int_{M}\cF = i(\omega ')$ for $\omega '=\int_{M}\corr
M\wedge \omega $.

Notice that
$\cE\bul(\pt)$~is {\it not\/} concentrated in degree zero.  For example, even
in ordinary cohomology the exact sequences ~\thetag{2.5} and~\thetag{2.6}
imply
  $$ \check H\bul(\pt) = \cases \ZZ
     ,&\bullet=0,\\\TT,&\bullet=1,\\0,&\text{otherwise}.\endcases
      $$

A cohomology theory determines a Pontrjagin dual cohomology theory.  The
cohomology theories which express the quantization law for self-dual fields
are quite special: they are Pontrjagin self-dual.  See Appendix~B, especially
\theprotag{B.2} {Definition}, for a discussion.

 \subhead Self-dual abelian gauge fields 
 \endsubhead

We now describe a generalized self-dual field.  Just as a $\sigma $-model
depends on a target Riemannian manifold as well as a spacetime dimension, so
too does a generalized self-dual field depend on external data.  We begin by
specifying enough of this data to define the Hamiltonian theory, then write
the classical theory in a Lorentzian spacetime of the Hamiltonian
form~$\RR\times Y$, and finally define the quantum Hilbert space of a
self-dual field.  We emphasize that our definition of the quantum Hilbert
space and certain operators on it is only up to {\it noncanonical\/}
isomorphism.  This is enough to demonstrate the Heisenberg uncertainty
principle for flux.  Thus the following definition is only good enough for
our noncanonical construction here; we need more precise data for the full
quantum theory.

        \definition{\protag{2.9} {Definition}}
   The data which define a {\it Hamiltonian self-dual generalized abelian
gauge field\/} are:\newline
 \indent(i)\ a Pontrjagin self-dual multiplicative cohomology
theory~$(E^\bullet,i)$ with shift~$s\in \ZZ$;\newline  
  \indent(ii)\ a dimension~$m$ and a multi-degree~$\dd$, i.e., an
ordered collection $\dd=(d_1,d_2,\dots ,d_k)$ of integer degrees;\newline
  \indent and (iii) a natural isomorphism
  $$ \phi \: E^{\dd}\longrightarrow E^{m-s+1-\dd} \tag{2.10} $$
such that for any compact $\cE$-oriented manifold~$Y$ of dimension~$m$ the
pairing
  $$ \aligned
      s_Y\: \cE^{\dd}(Y) \times \cE^{\dd}(Y)&\longrightarrow \TT \\
      (\bF_1, \bF_2) &\longmapsto \crho \int_{Y}\cphi (\bF_1)\cdot
     \bF_2\endaligned \tag{2.11} $$
is skew-symmetric.
        \enddefinition

\flushpar
 In~(i) $s\in \ZZ$ is a shift in degree---part of Pontrjagin
self-duality---and $i$~determines a homomorphism $E^{-s}(\pt)\to\ZZ$.  (In
all cases we know $i$~is specified by that homomorphism.)  In~(ii) $m$~is the
dimension of space, so we are working in a theory with $(m+1)$-dimensional
spacetimes.  The field strength of the self-dual field has degree~$\dd$,
which is a multi-degree in case there are several gauge fields.  The notation
for multi-degrees is $\dd+1=(d_1+1,d_2+1,\dots ,d_k+1)$, etc.  The physical
meaning of~(iii) is that $\phi $~induces a map of a non-self-dual gauge field
to its dual; it appears in the classical equation of motion~\thetag{2.12}
satisfied by a self-dual gauge field.  See~\theprotag{B.6} {Proposition} for
a description of the pairing in~\thetag{2.11}.  The self-dual scalar is the
case when $E$~is ordinary cohomology, $m=1$, $d=1$, and $\phi $~is the
identity map.  Other examples are given in the next subsection.  We remark
that whereas the pairing~\thetag{2.11} is skew-symmetric, it is not
necessarily alternating---that is, its values on the diagonal may be
nontrivial.  It is always a perfect pairing by \theprotag{B.6}
{Proposition}.

The Maxwell story of~\S{1} generalizes to self-dual gauge fields.
(\theprotag{2.18} {Example} below explains how to write Maxwell theory as a
special case.)  Let $Y$~be an $E$-oriented Riemannian manifold of
dimension~$m$ and~$M=\RR\times Y$ the associated Lorentzian spacetime of
signature~$(1,n-1)$.  The role of the electromagnetic field is played by a
differential form of total degree~$\dd$ on spacetime:
  $$ F\in \Omega (M;\VE)^{\dd}=\bigoplus\limits_{q\in \ZZ} \,\Omega
     ^q(M;\VE^{\dd-q}).  $$
The analog of the {\it classical\/} Maxwell equations~\thetag{1.1} with zero
current are the self-duality equations
  $$ \aligned
      dF &=0 \\
      \phi (F) &= \crho (*F)\endaligned \tag{2.12} $$
The second equation, the self-duality condition, takes values in~$\Omega
(M;\VE)^{m-s+1-\dd}$.  For the notation observe first that
\thetag{2.10}~applied to~$\RR^q$ with compact supports gives, after tensoring
with~$\RR$, a map $\VE^{\dd-q}\to\VE^{m-s+1-\dd-q}$.  Thus on the component
of~$F$ which is a differential form of degree~$q$ we have
  $$ \phi \:\Omega ^q(M;\VE^{\dd-q})\longrightarrow  \Omega
     ^{q}(M;\VE^{m-s+1-\dd-q}).  $$
On the right hand side appears the Lorentzian Hodge star operator 
  $$ *\:\Omega ^{m+1-q}(M;\VE^{\dd+q-m-1})\longrightarrow \Omega
     ^{q}\bigl(M;(\VE^{\dd+q-m-1})^*\bigr)  $$
followed by the duality map~\thetag{B.10}
  $$ \crho\:\Omega ^{q}\bigl(M;(\VE^{\dd+q-m-1})^*\bigr)\longrightarrow
     \Omega ^{q}\bigl(M;\VE^{m-s+1-\dd-q}\bigr). \tag{2.13} $$
We remark that if there is a twisting in the definition of the gauge field,
then the de Rham differential in the first equation is also twisted; see
\theprotag{2.27} {Example}.  Write $F=B-dt\wedge E$ as in~\thetag{1.3}; then
\thetag{2.12}~implies
  $$ \phi (B) = \crho(\star E),  $$
where $\star$~is the Riemannian Hodge star operator on~$Y$.  In other words,
up to some invertible algebraic maps the self-duality equates the electric
field and the magnetic field.  The classical flux, now both electric and
magnetic combined, is an $H(Y;\VE)^{\dd}$-valued function on the space of
solutions to~\thetag{2.12}, defined simply as the de Rham cohomology class
of~$F$.  There is a Poisson structure with symplectic leaves parametrized
by~$H(Y;\VE)^{\dd}$ and the classical fluxes Poisson commute.  The
self-duality equation~\thetag{2.12} is a first-order linear hyperbolic
equation, and so a solution is determined by the value at any fixed time, as
for the special case of the self-dual scalar field in two
dimensions---see~\thetag{2.1}.  Specifically, the space of solutions is
isomorphic to the real vector space~$\Omega (Y;\VE)^{\dd}_{\text{closed}}$.

In the semiclassical picture with Dirac charge quantization the field is a
geometric representative of a class in~$\cE^{\dd}(M)$.  The space of
classical solutions on~$M$ is the differential cohomology
group~$\cE^{\dd}(Y)$.  Our definition of its quantization is motivated by the
discussion at the beginning of~\S{2} of the self-dual scalar in two
dimensions, as well as computations in other examples.  The
pairing~\thetag{2.11} is skew bimultiplicative, but not necessarily
alternating.  We apply \theprotag{A.3} {Proposition} of Appendix~A to
construct the corresponding graded Heisenberg group.

        \proclaim{\protag{2.14} {Theorem}}
 There exists a central extension
  $$ 1 \longrightarrow \TT\longrightarrow \sG_Y\longrightarrow
     \cE^{\dd}(Y)\longrightarrow 0  $$
of $\agp=\cE^{\dd}(Y)$, unique up to noncanonical isomorphism, with graded
commutator~\thetag{2.11}.
        \endproclaim

\flushpar 
 Then \theprotag{A.5} {Proposition} and the remarks which follow imply

        \proclaim{\protag{2.15} {Theorem}}
 There exists an irreducible $\zt$-graded unitary
representation~$\sH=\sH^0\oplus \sH^1$ of~$\sG_Y$ on which the central circle
acts by scalar multiplication, and it is unique up to noncanonical
isomorphism.
        \endproclaim

\flushpar
 The quantum Hilbert space of the self-dual field on~$Y$ is the irreducible
representation~$\sHY$ in \theprotag{2.15} {Theorem}.  The generalization of
\theprotag{1.20} {Theorem} to self-dual gauge fields is now immediate.  The
quantum fluxes are defined using the kernel torus in~$E^{\dd}(Y)$, lifted to
the Heisenberg group~$\sG_Y$.  Namely, for~$\omega \in
\ET^{\dd-1}(Y)\subset \cE^{\dd}(Y)$ and $\tom$~a lift to~$\sG_Y$, define
$\Fqtm(\tom )$ to be the corresponding unitary operator on~$\sHY$.  Then the
commutation relation in~$\sG_Y$ implies
  $$ [\Fqtm(\tom _1),\Fqtm(\tom _2)] = \stY(\omega _1,\omega
     _2)\,\id_{\sHY},\qquad \omega _i\in \ET^{\dd-1}(Y), \tag{2.16} $$
where $\stY$~is the pairing~\thetag{2.11} and the left hand side is the
graded {\it group\/} commutator~\thetag{A.2}.  The nondegeneracy of~$\stY$
implies the following.

        \proclaim{\protag{2.17} {Generalized Main Observation}}
 If $\Tors E^{\dd}(Y)\not= 0$, then not all fluxes commute.
        \endproclaim

\flushpar 
 This is the generalization of \theprotag{1.18} {Main Observation}.
Analogous to \theprotag{1.19} {Amelioration} there is a grading of~$\sHY$ by
$E^{\dd}(Y)/\tors$.

 \subhead Examples
 \endsubhead

The lagrangian versions of the following examples are discussed
in~\cite{F,\S3}.  The first example demonstrates how the framework of
self-dual fields encompasses ordinary gauge fields.

        \example{\protag{2.18 (Dual gauge fields)} {Example}}
 Fix a dimension~$m$ for space and a degree~$1\le d \le m$ for a gauge
field.  Set~$\dd=(d,m+1-d)$ and quantize using ordinary cohomology~$E=H$.
The natural isomorphism~$\phi $ in~\thetag{2.10} is
  $$ \phi (\cF,\cF') = \bigl((-1)^{d(m-d)} \cF',(-1)^{d-1}\cF\bigr),\qquad
     \cF\in \cH^d(Y),\quad \cF'\in \cH^{m+1-d}(Y), \tag{2.19} $$
for any compact oriented manifold~$Y$.  The homomorphism~$i\:H^0(\pt)\to\ZZ$
is the natural isomorphism.  (The signs in equation~\thetag{2.19} are derived
from~\cite{F,Example~3.17}.)
 
The Maxwell theory of~\S{1} is the case~$m=3$, $d=2$.  Then up to an overall
sign \thetag{2.11}~is equal to the commutator in that
theory---see~\thetag{1.25}, \thetag{1.27}, and \thetag{1.28}.  (Note that
\thetag{1.28}~is the product in~$\cH^2$.)  In the discussion of~\S{1} the
alternating form~$\stY$ is written as the skew-symmetrization of a
2-cocycle~$\psi $.  For a general self-dual field the 2-cocycle~$\psi $ is
not determined by the data, but rather its existence is guaranteed by
\theprotag{2.14} {Theorem}.  The ``lagrangian splitting''~\thetag{1.26} gives
rise to~$\psi $ as well as to a partitioning of fluxes into magnetic and
electric.
        \endexample

        \example{\protag{2.20 (Standard self-dual gauge field~\cite{HS},
\cite{W})} 
{Example}}
 Here $m=4k+1$ for an integer~$k\ge0$, we use ordinary
cohomology~$E=H$, and there is a single degree~$d=2k+1$.  The
automorphism~$\phi $ is trivial in this case.  The self-dual scalar in two
dimensions is the case~$k=0$.  Let $L^3$~be a three-dimensional lens space.
For~$k\ge 1$ the manifold $Y=L^3\times S^{2k-1}\times S^{2k-1}$ exhibits
noncommuting fluxes.
        \endexample

        \example{\protag{2.21 (Type II Ramond-Ramond field ($\cB=0$))}
{Example}} 
 Here $m=9$ and we quantize using complex $K$-theory~$E=K$.  We assume the
$B$-field vanishes.  The Ramond-Ramond field on a compact Riemannian spin
manifold~$Y$ has an equivalence class in~$\cK^d(Y)$, where $d=0$~in Type~IIA
and $d=-1$~in Type~IIB.  Recall that $K\bul(\pt)\cong \ZZ[u,u\inv ]$ is a
Laurent series ring with $\deg u=2$.  The automorphism~$\phi $ is essentially
complex conjugation:\footnote{Complex conjugation maps the inverse Bott
element~$u$ to~$-u$.}
  $$ \phi (\cF)= u^{\ell }\,\overline{\cF} ,\qquad \ell =\cases 5,&\text{Type
     IIA};\\6,&\text{Type IIB}\endcases. \tag{2.22} $$
The homomorphism $i\:K^0(\pt)\to\ZZ$ is the augmentation, an isomorphism.  To
check that \thetag{2.11}~is skew-symmetric in Type~IIA we note that for
any~$\cF_1,\cF_2\in \cK^0(Y)$, the element $\cF_1\overline{\cF_2} +
\cF_2\overline{\cF_1}$ is the complexification of the realification
of~$\cF_1\overline{\cF_2}$.   Using the periodicity of
$K$-theory we can shift the integral in~\thetag{2.11} to the right-hand
vertical map in the commutative diagram
  $$ \CD
      \cK^0(Y) @>>> \check{K}O^0(Y) @>>> \cK^0(Y)\\
      @VVV @VVV @VVV\\
       \cK^{-9}(\pt) @>>> \check{K}O^{-9}(\pt) @>>> \cK^{-9}(\pt) \endCD  $$
Starting with $\cF_1\overline{\cF_2}$ in the upper left-hand corner we deduce
the skew-symmetry, since the composition $\TT \to\zt\to\TT$ on the bottom
line is zero.  A similar argument works for Type~IIB.

We make the classical self-dual equations~\thetag{2.12} explicit for
Type~IIA.  The field strength~$F\in \Omega \bigl(\RR\times Y[u,u\inv
]\bigr)^0$ has an expansion 
  $$ F = F_0 + F_2u\inv + F_4 u^{-2} + F_6u^{-3} + F_8u^{-4} + F_{10}u^{-5},
     \tag{2.23} $$
where the subscript~$q$ indicate the degree of the scalar-valued differential
form~$F_q$, and the forms~$F_q$ are closed.  Then
  $$ \phi (F) = F_0 u^5 - F_2u^4 + F_4 u^3 - F_6 u^2 + F_8 u - F_{10}
     \tag{2.24} $$ 
and 
  $$ \crho (*F) = *F_0 + *F_2 u + *F_4u^2 + *F_6u^3 + *F_8u^4 +
     *F_{10}u^5. \tag{2.25} $$ 
The self-duality equation is then
  $$ F_6 = -*F_4,\qquad \qquad F_8 = *F_2,\qquad \qquad F_{10} =
     -*F_0. \tag{2.26} $$

In the quantum theory the \theprotag{2.17} {Generalized Main Observation}
asserts that any manifold with torsion in its $K$-theory exhibits
noncommuting fluxes.  For~$K^0$, hence for Type~IIA, there is torsion in
$K$-theory on any manifold with finite abelian fundamental group.
(Calabi-Yau manifolds with this property have been considered in the physics
literature.)  For~$K^1$, hence for Type~IIB, one can take the product of a
circle and a manifold with finite abelian fundamental group to obtain
examples.
	\endexample

        \example{\protag{2.27 (Type II Ramond-Ramond field ($\cB\not= 0$))}
{Example}} 
 Now we allow nonzero $B$-field, which is a cocycle~$\cB$ representing an
element of~$\cH^3(Y)$.  Then the Ramond-Ramond field has an equivalence class
in differential $K$-theory {\it twisted\/} by~$\cB$, denoted~$\cK^{d + \cB}$.
As before $d=0$~in Type~IIA and $d=-1$~in Type~IIB.  The
automorphism~\thetag{2.22} maps to differential $K$-theory twisted by~$-\cB$.
The de Rham differential in \thetag{2.12} is replaced by the ``twisted''
differential $d+u\inv H$, where $H\in \Omega ^3(Y)_{\text{closed}}$~is the
field strength of~$\cB$.  Equations~\thetag{2.23}--\thetag{2.26} hold as in
the untwisted case, but now the condition that $F$~be closed is replaced by
the equation $(d+u\inv H)F=0$.  In components this reads
  $$ dF_0=0,\quad \quad dF_2 + H\wedge F_0=0,\quad \quad dF_4 + H\wedge
     F_2=0,\quad \quad \text{etc.}  $$
	\endexample

        \example{\protag{2.28 (Type I Ramond-Ramond field)} {Example}}
The Type~I Ramond-Ramond field, or `$B$-field', is quantized by periodic real
$K$-theory, whose Pontrjagin self-duality is proved in \theprotag{B.11}
{Proposition}.  The degree of the field strength is~$d=-1$, as in Type~IIB.
Note the nonzero shift~$s=4$.  The ring~$KO^{\bullet}(\pt)$ has torsion;
after tensoring over the reals we obtain $V_{KO}\cong \RR[u^2,u^{-2}]$, where
$\deg u=2$ as in $K$-theory.  The automorphism~\thetag{2.10} is
  $$ \phi (\cF) = \lambda ^{-1}\,\cF,  $$
where $\lambda\inv  \in KO^{8}(\pt)$ is the generator whose complexification
is~$u^{4}\in K^{8}(\pt)$.  In fact, there is a background self-dual
current---part of the Green-Schwarz mechanism---and the $B$-field is a
`cochain' of degree~$-1$ which trivializes the current.  See~\cite{F} for
details on the lagrangian theory.\footnote{The Pontrjagin self-duality was
expressed in that paper in terms of a hybrid of real and quaternionic
$K$-theory, but should have been stated as it is here.}
        \endexample

 \head
 Appendix A: Central Extensions of Abelian Groups
 \endhead
 \comment
 lasteqno A@  5
 \endcomment

We shall consider the class of abelian Lie groups $\agp$ which fit into an
exact sequence
  $$ 0\longrightarrow \pi_1(\agp) \longrightarrow V \longrightarrow \agp \longrightarrow \pi_0(\agp) \longrightarrow 0, $$
where $V$ --- the Lie algebra of $\agp$ --- is a locally convex and complete
topological vector space. We shall assume that the exponential map $V \to \agp$
is a local diffeomorphism which makes $V$ a covering space of the connected
component of $\agp$, and that the group $\pi_0(\agp)$ of connected components,
and the fundamental group $\pi_1(\agp)$, are finitely generated discrete
abelian groups.

Requiring $\pi_0$ and $\pi_1$ to be finitely generated may seem unduly
restrictive, but it includes the examples we are interested in, and going
beyond it makes things decidedly more complicated. In particular, if $F$ is a
discrete closed subgroup of a topological vector space $V$ it need not be
true that a homomorphism $F \to \RR$ extends to a continuous homomorphism $V
\to \RR$ if $F$ is not finitely generated; for example, if $F$ is the subgroup
of a Hilbert space $V$ generated by the elements of an orthonormal basis
$\{e_n\}$ then the map which takes each $e_n$ to 1 does not extend
continuously to $V$.

We shall now classify the central extensions $\sG$ of $\agp$ by the circle
group $\TT$ --- we shall call these groups {\it generalized Heisenberg
groups}. In fact we shall describe the {\it category} $\sE_{\agp}$ of
extensions, in which a morphism from $\sG$ to $\sG'$ is an isomorphism $\sG
\to \sG'$ which makes the diagram
  $$ \xymatrix{\TT \ar[r] \ar@{=}[d] &\sG \ar[r] \ar[d] &\agp\ar@{=}[d]\\ \TT
     \ar[r]& \sG' \ar[r] & \agp } $$
commute. The class of extensions we consider are those which as manifolds are
smooth locally trivial circle bundles over $\agp$.  It follows from
\theprotag{A.1(ii)} {Proposition} that these circle bundles are in fact
globally trivial.

A smooth bimultiplicative map $$ \psi : \agp \times \agp \longrightarrow \TT
$$ defines an extension $\sG(\agp,\psi)$: as a manifold $\sG(\agp, \psi)$ is
the product $\agp \times \TT$, and its multiplication is given by $$
(a_1,\lambda_1) \, (a_2,\lambda_2) \ = \ (a_1 +a_2, \lambda_1 \lambda_2
\psi(a_1,a_2)). $$ (Notice that a bimultiplicative map automatically
satisfies the cocycle condition~\thetag{1.24}.)

Let us recall that a {\it quadratic} map $f:\agp \to \TT$ is a smooth map
such that $$ (a_1,a_2) \mapsto \psi_f(a_1,a_2) = f(a_1 +
a_2)f(a_1)^{-1}f(a_2)^{-1} $$ is bimultiplicative. We introduce the category
$\sB_{\agp}$ whose objects are the smooth bimultiplicative maps, and whose
morphisms from $\psi$ to $\psi '$ are the smooth quadratic maps $f$ such that
$\psi_f = \psi'/\psi.$ There is a functor $\sB_{\agp} \to \sE_{\agp}$, as
a quadratic map $f$ defines a homomorphism $\sG(\agp,\psi) \to \sG(\agp,\psi
\psi_f)$ by
  $$(a,\lambda) \mapsto (a,\lambda f(a)).$$

        \proclaim{\protag{A.1} {Proposition}}
 \rom(i\rom)\ An object of $\sE_{\agp}$ is determined up to isomorphism by
its commutator map
  $$ s: \agp \times \agp \longrightarrow \TT, $$  
which is a bimultiplicative alternating map, and every bimultiplicative
alternating map can arise.\newline 
 \rom(ii\rom)\ The functor $\sB_{\agp} \to \sE_{\agp}$ is an equivalence of
categories.
        \endproclaim

        \demo{Proof}
 To prove that an extension is determined by its commutator map we observe
that the category $\sE_{\agp}$ has a composition-law. For the fibre-product
$\sG_1 \times_{\agp} \sG_2$ over $\agp$ of two extensions $\sG_1$ and $\sG_2$
is an extension of $\agp$ by $\TT \times \TT$, and we can define the desired
composite $\sG_1 * \sG_2$ as the quotient of the fibre-product by the
antidiagonal subgroup of $\TT \times \TT$ --- the image of $\TT$ by $\lambda
\mapsto (\lambda, \lambda^{-1})$. The isomorphism-classes of $\sE_{\agp}$
form a group Ext$(\agp)$ under this composition-law, for the inverse of $\sG$
is the quotient of the product extension $\sG \times \TT$ of $\agp$ by $\TT
\times \TT$ by the diagonal subgroup of $\TT \times \TT$. Using the
composition-law, it is enough to show that an extension which is itself
commutative is trivial. But any of the abelian groups of the type we are
considering can be written non-canonically as a product $K \times V \times
\pi$, where $K$ is a compact torus, $V$ is a vector space, and $\pi$ is
discrete. (To see this, we first split off $\pi$, using the fact that the
identity component of the group is divisible; then we split the Lie algebra
of the identity-component as $V$ times the Lie algebra of the
finite-dimensional maximal compact subgroup $K$, using the version of the
Hahn-Banach theorem which asserts that any continuous linear map $V_0 \to
\RR$ defined on a closed subspace of a locally convex topological vector
space $V$ can be extended to a continuous linear map $V \to \RR$.) This
reduces us to showing that any inclusion of a circle in a torus is split,
which follows from Pontrjagin duality.

To complete the proof of \theprotag{A.1} {Proposition}, let Bim$(\agp_1,
\agp_2)$ denote the group of bimultiplicative maps $\agp_1 \times \agp_2 \to
\TT$, and let Alt$(\agp)$ denote the group of alternating maps $\agp \times
\agp \to \TT$. There is an obvious map Bim$(\agp, \agp) \to$ Alt$(\agp)$
which takes $\psi$ to the commutator $$ (a_1,a_2) \mapsto s(a_1,a_2) =
\psi(a_1,a_2)/\psi(a_2,a_1) $$ of the extension $\sG(\agp,\psi)$. We need
only show that Bim$(\agp,\agp) \to$ Alt$(\agp)$ is surjective for all abelian
groups $\agp$ in our class. Using the fact that $$ {\Alt}(\agp_1 \oplus
\agp_2) \cong {\Alt}(\agp_1) \oplus {\Alt}(\agp_2) \oplus
{\Bim}(\agp_1,\agp_2) ,$$ together with the more obvious decomposition of
${\Bim}(\agp_1 \oplus \agp_2,\agp_1 \oplus \agp_2)$ as a sum of four groups,
we see that if the desired surjectivity holds for $\agp_1$ and $\agp_2$ then
it holds for $\agp_1 \oplus \agp_2$. It is therefore enough to consider
separately the cases when $\agp$ is a torus, a topological vector space, and
a discrete cyclic group, and each of these is trivial.
        \enddemo

In fact the abelian groups $\agp$ which are important in the present work
come equipped with skew bimultiplicative maps $s:\agp \times \agp \to \TT$
rather than alternating ones --- i.e. $s(b,a) = s(a,b)^{-1}$, and so $s(a,a)$
has order 2, but we need not have $s(a,a) = 1$. Such a skew map $s$ defines a
mod 2 grading $\agp = \agp^{\text{even}} \cup \agp^{\text{odd}}$ of $\agp$,
as $a \mapsto s(a,a)$ is a continuous homomorphism $\agp \to \ZZ/2$. In fact
the group Skew$(\agp)$ of these forms can be identified with the group
Ext$_{\text{gr}}(\agp)$ of isomorphism classes of {\it graded central
extensions}, as we shall now explain.

A central extension of a --- not necessarily abelian --- group $\agp$ by
$\TT$ can be regarded as a rule that associates a hermitian line $L_a$ to
each $a \in \agp$, and associative unitary isomorphisms $m_{a,b}: L_a \otimes
L_b \to L_{a+b}$ to each $a,b \in \agp$. (The elements of the extended group
are then pairs $(a,\lambda)$, with $a \in \agp$ and $\lambda \in L_a$ of unit
length.)  To define a graded central extension we simply replace the lines
$L_a$ by {\it graded} lines (with a mod 2 grading), and require the
isomorphisms $m_{a,b}$ to preserve the grading. A graded line is either even
or odd, and evidently the group $\agp$ acquires a grading from the degree of
$L_a$. In fact a graded central extension is simply a central extension $\sG$
of $\agp$ by $\TT$ which is at the same time a mod 2 graded group --- i.e. is
equipped with a homomorphism $\sG \to \ZZ/2$.

The analogue of the commutator map $s:\agp \times \agp \to \TT$ in the graded
case is the graded commutator, which maps $(a,b) \in \agp \times \agp$ to the
composite  
  $$ s(a,b) = m_{b,a} \circ T_{a,b} \circ m_{a,b}^{-1} :L_{a+b}
     \longrightarrow L_{a+b}, \tag{A.2} $$
where $T_{a,b}: L_a \otimes L_b \to L_b \otimes L_a$ expresses the symmetry
of the graded tensor product --- i.e. it multiplies by $-1$ if both $L_a$ and
$L_b$ are odd. We think of $s(a,b)$ as an element of $\TT$: concretely, it is
$$ (-1)^{\deg (L_a) \deg (L_b)} $$ times the naive commutator in the
extension.

Graded central extensions of $\agp$ possess a composition law (which takes
$\{L_a\}$ and $\{L'_a\}$ to $\{L_a \otimes L'_a\}$), and there is an obvious
short exact sequence $$ 0 \longrightarrow {\text{Ext}}(\agp) \longrightarrow {\text{Ext}}_{\text{gr}}(\agp) \longrightarrow
{\text{Hom}}(\agp ; \ZZ/2) \longrightarrow 0. $$ Comparing this with the other obvious exact
sequence $$ 0 \longrightarrow {\text{Alt}}(\agp) \longrightarrow {\text{Skew}}(\agp) \longrightarrow {\text{Hom}}(\agp ; \ZZ/2)
\longrightarrow 0, $$ we have

        \proclaim{\protag{A.3} {Proposition}}
 An object of $\sE^{\text{gr}}_{\agp}$ is determined up to isomorphism by its
graded commutator map
  $$ s: \agp \times \agp \longrightarrow \TT, $$  
which is a skew bimultiplicative map, and every skew bimultiplicative map can
arise.
        \endproclaim

We can also make an assertion analogous to \theprotag{A.1(ii)} {Proposition}
in the graded case: if we consider the category of graded extensions with a
{\it given} grading on $\agp$ the assertion is verbally the same as in the
ungraded case.

 \subhead Representations
 \endsubhead

Our next task is to describe the irreducible unitary representations of a
generalized Heisenberg group $\sG(\agp,\psi)$. It will be enough to consider
representations in which the central subgroup $\TT$ acts by multiplication:
the general case reduces easily to this one.

The centre $\sZ$ of the group $\sG = \sG(\agp,\psi)$ is an extension of $Z$
by $\TT$, where $Z$ is the kernel $\{a \in \agp : s(a,b)= 1 \ {\text{for \
all}} \ b \in \agp \}$ of the commutator map. Being abelian, the extension
$\sZ$ is split, but not canonically, for the cocycle $\psi$ need not vanish
on $Z$. By Schur's lemma in any irreducible representation $\rho$ of $\sG$
the subgroup~$\sZ$ acts by scalar multiplication, i.e. by a homomorphism
$\chi:\sZ \to \TT$ which is a splitting of the extension.

        \proclaim{\protag{A.4} {Proposition}}
 For a finite-dimensional generalized Heisenberg group $\sG$ an irreducible
unitary representation $\rho$ is determined up to isomorphism by the
splitting homomorphism $\chi:\sZ \to \TT$, and any such homomorphism can
arise.
        \endproclaim

        \demo{Proof}
 If the Heisenberg group $\sG$ is non-degenerate,
i.e. its centre is exactly $\TT$, the assertion is that there is only one
possible representation. This is essentially the classical theorem of Stone
and von Neumann. Let us assume that it is known in the case when $\agp$ is a
finite-dimensional vector space. In general, let $T$ be the kernel of the
commutator map restricted to the identity component of $\agp$; this is
necessarily a torus over which the extension $\sG$ is canonically split (for
the bimultiplicative cocycle $\psi$ which defines $\sG$ must vanish when
restricted to a torus). Furthermore, $T$ is Pontrjagin-dual to $\sG/\sG_1$,
where $\sG_1$ is the centralizer of $T$ in $\sG$. Any Hilbert space $\sH$ on
which $\sG$ acts unitarily can be decomposed according to the action of $T$ as
$\sH = \oplus \sH_{\alpha}$, where $\alpha$ runs through the characters of
$T$. Now $\sH_0$ is a representation of $\sG_1$, and evidently $\sH$ is the
representation of $\sG$ induced from the representation $\sH_0$ of $\sG_1$,
which must therefore be irreducible. In fact $\sH_0$ is a representation of
the non-degenerate Heisenberg group $\sG_1/T$, whose group of components is
finite, and whose identity-component is a vector Heisenberg group. We have
therefore reduced ourselves to treating the case of a Heisenberg group $\sG$
arising from an abelian group $\agp$ of the form $V \times \pi$, where $V$ is a
vector space and $\pi$ is a finite abelian group. In this case, let $\Gamma$
be a maximal abelian subgroup of the restriction of the extension to
$\pi$. We can write $\Gamma = \TT \times \pi'$, though non-canonically. In any
representation of $\sG$ the action of the subgroup $\pi'$ decomposes the
Hilbert space into pieces according to the characters of $\pi'$, which are
permuted transitively by the conjugation-action of $\pi$. By the argument we
have already used we see that the representation is induced from an
irreducible representation of $\sG_0 \times \pi'$ in which $\pi'$ acts
trivially. (Here $\sG_0$ is the identity-component of $\sG$.) But $\sG_0$ is a
vector Heisenberg group, and so we have reduced ourselves to the case we have
assumed to be known.

When the group $\sG$ is degenerate, the representation $\rho$ is actually a
representation of $\sG/Z'$, where $Z'$ is the kernel of the homomorphism
$\chi:\sZ \to \TT$. But $\sG/Z'$ is a non-degenerate Heisenberg group,
and so we are back to the previous case.
        \enddemo

Turning now to infinite-dimensional groups, to have an analogue of the
Stone-von Neumann theorem we must introduce the concept of a {\it
positive-energy} representation, which is defined when the abelian group $\agp$
is {\it polarized}. There are many versions of this concept. The following is
a rather narrow one, but seems simplest for our purposes.

We shall say $\agp$ is polarized if there is a continuous action of the group
$\RR$ on the Lie algebra $V$ of $\agp$ by operators $\{u_t\}_{t \in \RR}$
which preserve the skew form coming from the commutator and decompose the
complexification $V_{\CC}$ into a countable sum of finite-dimensional
subspaces $V_{\lambda}$ in which $u_t$ acts by multiplication by $e^{i\lambda
t}$. Here each $\lambda$ is real, and we assume that the algebraic sum of the
$V_{\lambda}$ is dense in $V$. In our applications $\{u_t\}$ will be the
Hamiltonian flow on the phase space $V$ induced by a positive quadratic
energy-function --- but we allow the symplectic structure on $V$ to be
degenerate.

Then for a polarized group $\agp$ we say a unitary representation of the
Heisenberg group $\sG$ on a Hilbert space $\sH$ is of positive energy if there
is a unitary action of $\RR$ on $\sH$ by operators $U_t = e^{iHt}$
whose generator $H$ has discrete non-negative spectrum, and which intertwines
with the action of $\sG$ on $\sH$ in the sense that the action of exp$(u_t(v))$
for any $v \in V$ is the conjugate by $U_t$ of the action of exp$(v)$. In our
applications $\{U_t\}$ will be the time-evolution of a quantum system, which
we require to have positive energy in the usual quantum-mechanical sense.

Now we have the following version of the Stone-von Neumann theorem.

        \proclaim{\protag{A.5} {Proposition}}
 For a polarized generalized Heisenberg group $\sG$ an irreducible unitary
representation of positive energy is completely determined by the splitting
homomorphism $\chi:\sZ \to \TT$, and any such homomorphism can occur.
        \endproclaim

        \demo{Proof}
 By exactly the same arguments as in the finite-dimensional case we reduce
first to the case of a non-degenerate Heisenberg group, and then to a vector
Heisenberg group. We then find, just as in~\cite{PS,\S9.5}, that the unique
positive energy irreducible unitary representation of the Heisenberg group
formed from $V$ is realized on the completion of the symmetric algebra
$S(W)$, where $V_{\CC}= W \oplus \bar W$ is the decomposition into positive-
and negative-energy pieces.
        \enddemo

To conclude, we return to graded central extensions. For a mod 2 graded group
$\sG$ it is natural to consider unitary representations on mod 2 graded
Hilbert spaces, and to require that the action of the even elements of $\sG$
preserves the grading, while that of the odd elements reverses it. But if
$\sG$ is a graded generalized Heisenberg group every representation
automatically has such a grading. To see this, we may as well assume that
$\sG$ is non-degenerate, for in any irreducible representation $\sG$ acts
through a non-degenerate quotient. If $\sG$ is non-degenerate, the grading
homomorphism $\agp \to \ZZ/2$ is necessarily of the form $a \mapsto
s(\varepsilon, a)$ for some $\varepsilon \in \agp$ which has order 2. If
$\hat \varepsilon$ is a lift of $\varepsilon$ in $\sG$ then $\hat
\varepsilon$ commutes with the even elements of $\sG$ and anticommutes with
the odd elements. Furthermore, $\hat \varepsilon^2$ acts as a scalar, so its
eigenspaces define the desired mod 2 grading on the Hilbert space. (We do not
have a preferred way of naming the eigenspaces odd and even; but that does
not matter, as reversing the choice gives us an isomorphic graded
representation.)

 \head
 Appendix B: Self-Dual Cohomology Theories
 \endhead
 \comment
 lasteqno B@ 17
 \endcomment

Any cohomology theory $E^\bullet$ arises from a {\it loop-spectrum}, i.e. a
sequence $\sE = \{\sE_q\}_{q \in \ZZ}$ of spaces with base-point such that
$E^q(X)$ is the set of homotopy classes of maps $X \to \sE_q$.  For a space
$X$ with base-point, the reduced cohomology $\tilde E^q(X)$ is the set of
homotopy classes of base-point-preserving maps~$X\to\sE_q$.  The term
``loop"-spectrum reflects the existence of canonical homotopy equivalences
$\sE_q \to \Omega\sE_{q+1}$ --- where $\Omega$ denotes the based loop-space
--- which express the behaviour of cohomology groups under suspension. The
spectrum is determined up to homotopy equivalence by the cohomology theory,
and algebraic topologists usually mean the spectrum when they refer to a
cohomology theory. The spectrum also defines a homology theory $E_\bullet$ by
$$ E_q(X) = \lim _i \pi_{q+i}(X_+ \wedge \sE_i), $$ where $X_+$ denotes the
space $X$ --- which is not assumed to have a given base-point --- with a
disjoint base-point adjoined.\footnote{For two spaces $Y$ and $Z$ with
base-points $y_0$ and $z_0$ the wedge product $Y \wedge Z$ denotes the spaces
obtained from the product $Y \times Z$ by collapsing the subspace $(Y \times
z_0) \cup (y_0 \times Z)$ to a single point.}  \ In particular, $E_q(\pt ) =
E^{-q}(\pt )$.

It may be helpful to notice that the spaces $X_+ \wedge \sE_q$ form a {\it
spectrum}, in the sense that there are natural maps $$X_+ \wedge \sE_q \ \longrightarrow \
\Omega (X_+ \wedge \sE_{q+1}).$$ This is not a loop-spectrum, but one can make
it into a loop-spectrum $X_+ \otimes \sE$ whose $q$-th space is $$ \lim _i
\Omega^i(X_+ \wedge \sE_{q+i}), $$ and then $E_q(X) $ is the $q$-th homotopy
group of $X_+ \otimes \sE$.

In general there is no way to calculate the {\it groups} $E_\bullet(X)$
algebraically from the {\it groups} $E^\bullet(X)$, although if $X$ is compact and
can be embedded as a neighbourhood-deformation-retract in Euclidean space
$\RR^N$ we have
 $$E_q(X) \cong \tilde E^{N-q}({\roman D}^NX),$$
where D$^NX$ is the $N$-{\it dual} of $X$, defined as the one-point
compactification of an open neighbourhood of $X$ in $\RR^N$ of which $X$ is a
deformation-retract, and $\tilde E^\bullet$ denotes {\it reduced} cohomology.
This notion of duality --- usually called {\it S-duality} --- is explained by
the fact that the category of spectra forms --- when the morphisms are
defined appropriately --- a tensor category in which the tensor product is
induced by the wedge-product of spaces and the sphere-spectrum $\{S^q\}$ is
the neutral object. In this category the natural dual of a compact space $X$
is the spectrum formed by the mapping-spaces $\{ {\Map}(X;S^q)\}$. The
sequence of spaces $\{{\roman D}^qX\}$ forms a spectrum, and there is a
natural map --- the ``scanning" map\footnote{To define the scanning map,
choose $\varepsilon > 0$ so that the $\varepsilon$-neighbourhood of $X$ in
$\RR^q$ in contained in the open neighbourhood $U_X$ of $X$ of which ${\roman
D}^q(X)$ is the compactification. Then the scanning map is the extension to
the compactification of the map which takes $x \in U_X$ to the composite $U_X
\to \RR^q \to \RR^q/(\RR^q - U_x) \cong S^q$, where $U_x$ is the
$\varepsilon$-neighbourhood of $x$ in $\RR^q$.} --- ${\roman D}^qX \to
{\Map}(X;S^q)$ which induces an equivalence of the associated loop-spectra.

\bigskip

If $E^\bullet$ is a multiplicative theory, in the sense that $X \mapsto E^\bullet(X)$ is
a contravariant functor to anticommutative graded rings, then the
multiplication is induced by maps $\sE_p \wedge \sE_q \to \sE_{p+q}$. These
define maps
  $$ E^p(X) \times E_q(X) \longrightarrow E_{q-p}(X) $$
taking $f:X_+ \to \sE_p$ and $g:S^{q+i} \to X_+ \wedge \sE_i$ to the
composite
  $$ S^{q+i} \longrightarrow X_+ \wedge \sE_i \longrightarrow X_+ \wedge \sE_i \wedge \sE_p \longrightarrow X_+
     \wedge \sE_{p+i}, $$
which make $E_\bullet(X)$ a graded module over $E^\bullet(X)$.

\bigskip

There is another way --- essentially purely algebraic --- to pass between
cohomology theories and homology theories which from the point of view of
spectra does not look at all natural. Let us recall that if $I$ is a {\it
divisible} abelian group such as $\RR$ or $\TT$ then the contravariant functor
$A \mapsto {\Hom}(A;I)$ from the category of abelian groups to itself
takes exact sequences to exact sequences. So if $E_\bullet$ is a homology theory we
can define a cohomology theory $e^\bullet_I$ by
$$e^q_I(X) = {\Hom}(E_q(X);I).$$
Unfortunately this does not work when $I=\ZZ$, as $\ZZ$ is not a divisible
group. It is reasonable, however, to define the theory $e^\bullet =
e^\bullet_{\ZZ}$ as the theory that fits into a long exact sequence  
  $$ \ldots \longrightarrow e^q(X) \longrightarrow e^q_{\RR}(X)
     \longrightarrow e^q_{\TT}(X) \longrightarrow e^{q+1}(X) \longrightarrow
     \ldots \ , $$
where the transformation of theories $e^\bullet_{\RR} \to e^\bullet_{\TT}$ is
induced by the obvious homomorphism $\RR \to \TT$. (This does entail
representing $e^\bullet_{\RR}$ and $ e^\bullet_{\TT}$ by spectra, and
defining $e^\bullet$ as the theory represented by the fibre of the map of
spectra; but it is a fairly anodyne operation as we shall in the end have a
short exact sequence $$ 0 \longrightarrow {\Ext}(E_{q-1}(X);\ZZ)
\longrightarrow e^q(X) \longrightarrow {\Hom}(E_q(X); \ZZ) \longrightarrow
0.) $$

\bigskip

Because $E_\bullet(X)$ is a module over $E^\bullet(X)$ it is easy to see that
the cohomology theories $e^\bullet_I$ and $e^\bullet$ are module-theories
over $E^\bullet$.  Let $s$~be an integer termed a ``shift''.  If we choose
once and for all an element of $i \in e^s(\pt )$ --- which is simply
${\Hom}(E^{-s}(\pt ); \ZZ)$ if $E^{1-s}(\pt )=0$, as is the case for all the
theories we shall be concerned with --- then we get a natural element in $i_X
\in e^s(X)$ for all spaces $X$, and hence a natural transformation 
  $$ E^\bullet(X) \longrightarrow e^{\bullet+s}(X) \tag{B.1} $$
which takes $\xi$ to $\xi . i_X$.

        \definition{\protag{B.2} {Definition}}
 The theory $(E^\bullet,i)$ is called {\it Pontrjagin self-dual} if
\thetag{B.1}~defines an isomorphism of cohomology theories.
        \enddefinition

\flushpar
 This is a very strong constraint on a theory, as it implies, for instance,
that the groups $E^q(\pt )$ and $E^{-s-q}({\pt})$ have the same rank for
every $q$.  Known examples include classical homology and periodic complex
and real $K$-theory --- the shift~$s$ is zero for the former two and we take
$s=-4$ for the latter; see below.

In treating the examples it is more convenient to use an equivalent version
of the self-duality condition which is stated in terms of the theory
$E^\bullet_{\TT}$ called ``$E^\bullet$ with coefficients in $\TT$". This is a
module-theory over $E^\bullet$ which fits into a long exact sequence  
  $$ \ldots \longrightarrow E^q(X) \longrightarrow E^q_{\RR}(X)
     \longrightarrow E^q_{\TT}(X) \longrightarrow E^{q+1}(X) \longrightarrow
     \ldots\ , \tag{B.3} $$
where --- at least if certain finiteness conditions are satisfied which hold
in the cases at hand --- the theory $E^\bullet_{\RR}$ with real coefficients
is defined simply by $E^q_{\RR}(X) = E^q(X)\otimes \RR$.

The transformation $E^\bullet \to e^{\bullet+s}$ induced by a choice of $i$
also leads to a transformation $E^\bullet_{\TT} \to e^{\bullet+s}_{\TT}$, and
the latter is an isomorphism if and only if the former is. To check
self-duality, therefore, it is enough to prove that the map
$E^\bullet_{\TT}(\pt ) \to e^{\bullet+s}_{\TT}(\pt )$, which is a map of
modules over $E^\bullet(\pt )$, is an isomorphism, i.e. that the
module-action  
  $$ E^{q-s}(\pt ) \times E^{-q}_{\TT}(\pt ) \ \longrightarrow \
     E^{-s}_{\TT}(\pt ) \ @>\;\;i\;\;>> \ \TT \tag{B.4} $$
is a perfect pairing.

\bigskip

For a compact $E$-oriented $m$-manifold~$Y$ Poincar\'e duality there is a
perfect pairing
  $$ E^{m-s-q}(Y)\otimes E_{\TT}^{q}(Y)\longrightarrow \TT \tag{B.5} $$
for each integer~$q$.  For Pontrjagin self-duality identifies
$E_\TT^{q}(Y)\cong \Hom\bigl(E_{q+s}(Y);\TT \bigr)$ under which
\thetag{B.5}~becomes the composition
  $$ E^{m-s-q}(Y) \otimes \Hom\bigl(E_{q+s}(Y);\TT \bigr)\longrightarrow
     \Hom\bigl(E_{m}(Y);\TT \bigr)\longrightarrow \TT  $$
of the module-action of~$E^{\bullet}$ on~$e_\TT^{\bullet}$ and evaluation on
the fundamental class, and this is a perfect pairing by Poincar\'e duality.

Now let us turn to the differential theory $\check E^\bullet$ associated to
$E^\bullet$. If $E^\bullet$ is Pontrjagin self-dual then we have the
following very attractive differential version of Poincar\'e duality.  We
make the finiteness assumption alluded to after~\thetag{B.3}.

        \proclaim{\protag{B.6} {Proposition}}
If $Y$ is a compact m-dimensional manifold which is oriented for $\check
E^\bullet$, and the theory $(E^\bullet,i)$ is Pontrjagin self-dual, then we
have an integration operation
  $$ \crho\int_{Y}\:\check E^{m-s+1}(Y) \longrightarrow \TT \tag{B.7} $$
which, together with the natural multiplication, gives us a perfect pairing
  $$ \check E^q(Y) \times \check E^{m-s-q+1}(Y) \ \longrightarrow \
     \TT. \tag{B.8} $$

        \endproclaim

The proof relies on the following, which is also used in~\thetag{2.13}. 

        \proclaim{\protag{B.9} {Lemma}}
 For each~$q\in \ZZ$ the natural pairing
  $$  \crho\: \VE^{q-s}\otimes \VE^{-q}\to \RR \tag{B.10} $$
is nondegenerate. 
        \endproclaim

        \demo{Proof}
 By the finiteness assumption $\VE^q=E^q(\pt)\otimes \RR$.  The element~$i\in
e^s(\pt)$ induces a homomorphism $E^{-s}(\pt)\to\ZZ$, and so a linear map
$\crho\:\VE^{-s}\to\RR$ and the pairing~\thetag{B.10}.  If there is a nonzero
$v\in \VE^{q-s}$ in the kernel of the pairing, we can assume $v$~is in the
image of $E^{q-s}(\pt)\to E^{q-s}(\pt)\otimes \RR$, so defines an
element~$e\in E^{q-s}(\pt)/\text{torsion}$.  Now $E_{\RR}^{-q}(\pt)\to
E_{\TT}^{-q}(\pt)$ is onto the identity component, and the perfection of the
pairing~\thetag{B.4} implies that $E^{q-s}(\pt)/\text{torsion}\times
E_{\TT}^{-q}(\pt)_{\text{id}}\to\TT$ is also perfect, from which~$e=0$.
        \enddemo

        \demo{Proof of \theprotag{B.6} {Proposition}}
 First, \thetag{B.7}~is defined as the composition
  $$ \xymatrix@1{\cE^{m-s+1}(Y) \ar[r]^-{\tsize\int_{Y}} &\cE^{-s+1}(\pt)
     \ar@{=}[r]^-{\ssize\thetag{2.5}} & E^{-s}_{\TT}(\pt) \ar[r]^-i &\TT}
      $$
Suppose $\cF\in \cE^q(Y)$ is in the kernel of~\thetag{B.8} and let $F\in
\Omega _E(M;\VE)^q$ be its field strength.  Let~$\cG\in \cE^{m-s-q+1}(Y)$
have trivial characteristic class, so it is the image of a differential
form~$\alpha $ in the exact sequence~\thetag{2.6}.  Then by the remark
following~\thetag{2.7} and~\thetag{2.8} {\it et seq.\/} we see that $
\int_{Y}\cF\cdot \cG$ is image of $\int_{Y} \corr Y \wedge F\wedge \alpha$
under the natural map~$\RR\to\TT$.  Decompose~$F$ and~$\alpha $ as a sum of
differential forms of fixed degree:
  $$ \alignedat2
      F &=\sum\limits_{j}F _j,\qquad &&F _j\in \Omega
     ^j(\VE^{q-j}), \\
      \alpha &=\sum\limits_{k}\alpha _k,\qquad &&\alpha _k\in \Omega
     ^k(\VE^{m-s-q-k}) .\endaligned  $$
Then
since $\cF$~is assumed in the kernel of~\thetag{B.8}, and since $\corr Y$~is
invertible, it follows that $\crho$~applied to 
  $$ \int_{Y}F\wedge \alpha =\sum\limits_{j}\int_{Y}F _j\wedge \alpha
     _{m-j} \quad \in \quad \bigoplus\limits_j\;\VE^{q-j}\otimes \VE^{j-q-s}
      $$
vanishes for all~$\alpha $.  Then from \theprotag{B.9} {Lemma} we
conclude~$F=0$.  Thus $\cF$~is the image in~\thetag{2.5} of $\omega \in
E_{\TT}^{q-1}(Y)$, and furthermore $\omega $~lies in the kernel of the
pairing
  $$ E^{m-s-q+1}(Y)\otimes E_{\TT}^{q-1}(Y)\longrightarrow \TT.  $$
But this is the perfect Poincar\'e-Pontrjagin pairing~\thetag{B.5},
whence~$\omega =0$. 
        \enddemo

The Pontrjagin self-duality of ordinary cohomology and complex $K$-theory are
relatively easy, but for real $K$-theory it is more subtle.  In fact, the
Pontrjagin dual of $KO$-theory --- the associated theory denoted~$e^\bullet$
above --- is again $KO$-theory but with a shift of degree~$s=4$.  Note
$KO^{-3}(\pt)=0$ and $KO^{-4}(\pt)$~is generated by an element~$\mu $ whose
complexification is~$2u^{-2}\in K^{-4}(\pt)$.  We choose the element~$i\in
\Hom\bigl(KO^{-4}(\pt);\ZZ \bigr)$ to map~$\mu \mapsto 1$.

        \proclaim{\protag{B.11} {Proposition}}
 $(KO,i)$~is Pontrjagin self-dual.
        \endproclaim

 \midinsert
 \bigskip
 \centerline{\eightpoint
$$
\vbox{\offinterlineskip
\hrule
\halign{\vrule# &\quad\hfill #\hfill
  &\quad\vrule# &\quad\hfill #\hfill
  &\quad\vrule# &\quad\hfill #\hfill
  &\quad\vrule# &\quad\vrule# \cr
height6pt &\omit &&\omit &&\omit  &\cr
&$q$ &&$KO_q(\pt)\cong KO^{-q}(\pt)$
&&$KO^q_\TT(\pt)$ &\cr
height6pt &\omit  &&\omit &&\omit &\cr
\noalign{\hrule height 1.5pt depth 0pt}
height4pt &\omit   &&\omit &&\omit &\cr
&$4$ &&$\ZZ$\hbox to 0pt{\quad \ \ \  $(\mu )$}&&$\TT$  &\cr
height4pt &\omit  &&\omit &&\omit  &\cr
\noalign{\hrule}
height4pt &\omit   &&\omit &&\omit &\cr
&$3$ &&$0$&&$0$  &\cr
height4pt &\omit  &&\omit &&\omit  &\cr
\noalign{\hrule}
height4pt &\omit   &&\omit &&\omit &\cr
&$2$ &&$\zt$\hbox to 0pt{\quad $(\eta ^2)$}&&$0$  &\cr
height4pt &\omit  &&\omit &&\omit  &\cr
\noalign{\hrule}
height4pt &\omit   &&\omit &&\omit &\cr
&$1$ &&$\zt$\hbox to 0pt{\quad $(\eta )$}&&$0$  &\cr
height4pt &\omit  &&\omit &&\omit  &\cr
\noalign{\hrule}
height4pt &\omit   &&\omit &&\omit &\cr
&$0$ &&$\ZZ$&&$\TT$  &\cr
height4pt &\omit  &&\omit &&\omit  &\cr
\noalign{\hrule}
height4pt &\omit   &&\omit &&\omit &\cr
&$-1$ &&$0$&&$0$  &\cr
height4pt &\omit  &&\omit &&\omit  &\cr
\noalign{\hrule}
height4pt &\omit   &&\omit &&\omit &\cr
&$-2$ &&$0$&&$\zt$  &\cr
height4pt &\omit  &&\omit &&\omit  &\cr
\noalign{\hrule}
height4pt &\omit   &&\omit &&\omit &\cr
&$-3$ &&$0$&&$\zt$  &\cr
height4pt &\omit  &&\omit &&\omit  &\cr
\noalign{\hrule}
height4pt &\omit   &&\omit &&\omit &\cr
&$-4$ &&$\ZZ$&&$\TT$  &\cr
height4pt &\omit  &&\omit &&\omit  &\cr
\noalign{\hrule}
}
\hrule}
$$
}
 \nobreak
 \endinsert

        \demo{Proof}
 According to~\thetag{B.4} it suffices to verify that the pairing
  $$ KO^{q-4}(\pt) \otimes KO^{-q}_\TT(\pt)\longrightarrow
     KO_{\TT}^{-4}(\pt)@>{\;\  i\ \;}>>\TT  $$
is an isomorphism for all~$q$.  The chart above, together with Bott
periodicity $KO^{q+8}\cong KO^q$, reduces our task to the following four
statements:
  $$ \align
      KO^0(\pt)\otimes KO^{-4}_\TT(\pt)&\longrightarrow
     KO^{-4}_{\TT}(\pt)\text{ is an isomorphism;} \tag{B.12} \\
      KO^{-4}(\pt)\otimes KO^0_\TT(\pt)&\longrightarrow
     KO^{-4}_{\TT}(\pt)\text{ is an isomorphism;} \tag{B.13} \\
      KO^{-1}(\pt)\otimes KO^{-3}_\TT(\pt)&\longrightarrow
     KO^{-4}_{\TT}(\pt)\text{ is injective;} \tag{B.14} \\
      KO^{-2}(\pt)\otimes KO^{-2}_\TT(\pt)&\longrightarrow
     KO^{-4}_{\TT}(\pt)\text{ is injective.} \tag{B.15} \endalign $$
To verify~\thetag{B.12} and~\thetag{B.13} we introduce quaternionic
$K$-theory~$KSp$ and use Bott periodicity $ KSp^q\cong KO^{q+4}$.  Since
$KSp^0_\TT(\pt)\cong KSp^0(\pt)\otimes \TT$ and $KO^0_\TT(\pt)\cong
KO^0(\pt)\otimes \TT$, it suffices to know that the natural pairing
$KO^0(\pt) \otimes KSp^0(\pt)\to KSp^0(\pt)$ is an isomorphism, which is
clear: the generators of~$KO^0(\pt)$ and~$KSp^{0}(\pt)$ are the trivial real
and quaternionic lines, and the map is the tensor product over the reals.
      
Now the generator of~$KO^{-1}(\pt)\cong \zt$ is a class~$\eta $, and $\eta
^2$~generates~$KO^{-2}(\pt)$, whence \thetag{B.14}~follows from~\thetag{B.15}.
Let $KO^q_{\zt}(X)$~denote the $KO$-group with $\zt$-coefficients.  The
diagram of short exact sequences 
  $$ \xymatrix{0 \ar[r] &\ZZ \ar[r]^2 \ar@{=}[d] &\ZZ \ar[r]\ar[d]
     ^{1/2} &\zt\ar[r] \ar[d] &0 \\
      0 \ar[r] &\ZZ \ar[r] &\RR \ar[r] &\TT \ar[r] &0}  $$
induces a diagram of long exact sequences 
  $$ \xymatrix{\ar[r]^-{0} &KO^{-2}(\pt) \ar[r] \ar[d]^0 &KO^{-2}_{\zt}(\pt)
     \ar[r]\ar[d] &KO^{-1}(\pt) \ar[r]^-{0} \ar@{=}[d] &\\
      \ar[r] &KO^{-2}_\RR(\pt) \ar[r]^0 &KO^{-2}_{\TT}(\pt) \ar[r]
     &KO^{-1}(\pt) \ar[r]^-{0}& } \tag{B.16} $$
Each of $KO^{-1}(\pt)$~and $KO^{-2}(\pt)$~is cyclic of order two, from which
$KO^{-2}_{\zt}(\pt)$ is either isomorphic to~$\ZZ/4\ZZ$ or~$\zt\times \zt$;
we will see shortly that it is the former.  Then the sequence shows that the
generator of this group maps to the generator of~$KO^{-2}_\TT(\pt)\cong \zt$.
A similar argument with a different stretch of~\thetag{B.16} shows that
$KO^{-3}_{\zt}(\pt)$ is cyclic of order two, and with yet another stretch
of~\thetag{B.16} we find $KO^{-4}_{\zt}(\pt)$ is cyclic of order two and
$KO^{-4}_{\zt}(\pt)\to KO^{-4}_\TT(\pt)$ is injective.  Hence we are reduced
to the statement that $\eta ^2\:KO^{-2}_{\zt}(\pt)\to KO^{-4}_{\zt}(\pt)$ is
nonzero.  Cohomology groups with $\zt$~coefficients may be computed by
smashing with~$\RP^2$ and shifting degree by two,\footnote{For any cohomology
theory~$E$ and any pointed space~$X$,
  $$ E^q_{\zt}(X)\cong \tilde E^{q+2}(X_+\wedge \RP^2).  $$
} so we must show that 
  $$ \eta ^2\:\widetilde{KO}^0(\RP^2)\longrightarrow
     \widetilde{KO}^{-2}(\RP^2) \tag{B.17} $$
is nonzero.  Note that the former group is generated by $H-1$, where
$H\to\RP^2$~is the nontrivial real line bundle, and since $w_2(H\oplus
H)\not= 0$ it follows that $\widetilde{KO}^0(\RP^2)$ is cyclic of order four,
verifying the claim above.  Finally, we deduce that \thetag{B.17}~is nonzero
by applying the long exact ``Bott sequence''\footnote{One derivation begins
with the fibration $U/O \to BO \to BU$ and the Bott periodicity $\Omega
(U/O)\sim \ZZ\times BO $. }
  $$ \xymatrix@1{\ar[r] &\tilde{K}^{q-2}(\RP^2) \ar[r]
     &{\widetilde{KO}}^q(\RP^2) \ar[r]^-{\eta} &{\widetilde{KO}}^{q-1}(\RP^2)
     \ar[r] & \tilde{K}^{q-1}(\RP^2) \ar[r] &}  $$
twice.  First, set~$q=0$ and use $K^{-1}(\RP^2)=0$ to deduce that $\eta
\:\widetilde{KO}^0(\RP^2)\to\widetilde{KO}^{-1}(\RP^2)$ is surjective.  Then
set~$q=-1$ and use $K^{-3}(\RP^2)=0$ to deduce that $\eta
\:\widetilde{KO}^{-1}(\RP^2)\to\widetilde{KO}^{-2}(\RP^2)$ is an
isomorphism.  (Both groups are cyclic of order two.)
        \enddemo

\enddocument